\documentclass[prb,twocolumn,showpacs,amsmath,longbibliography,superscriptaddress,preprintnumbers,amssymb]{revtex4-1}
\usepackage{graphicx}% Include figure files
\usepackage{dcolumn}% Align table columns on decimal point
\usepackage{bm}% bold math
\usepackage{wasysym}
\usepackage{dsfont}
\usepackage{enumitem}
\newcommand{\beq}{\begin{equation}}
\newcommand{\eeq}{\end{equation}}
\newcommand{\beqn}{\begin{eqnarray}}
\newcommand{\eeqn}{\end{eqnarray}}

\newcommand{\cC}{ {\cal C} }
\newcommand{\cD}{ {\cal D} }

\newcommand{\cL}{ {\cal L} }

\newcommand{\cT}{ {\cal T}}

\newcommand{\ii}{\mathrm{i}}

\newcommand{\U}{\mathrm{U}}

\def\Z{\mathds{Z}}

\usepackage{xcolor}

\begin{document}

\title{Minimal Fractional Topological Insulator in half-filled conjugate moir\'{e} Chern bands}

\author{Chao-Ming Jian}
\affiliation{Department of Physics, Cornell University, Ithaca, New York 14853, USA}

\author{Meng Cheng}
\affiliation{Department of Physics, Yale University, New Haven, Connecticut 06511-8499, USA}

\author{Cenke Xu}
\affiliation{Department of Physics, University of California, Santa Barbara, CA 93106}

\begin{abstract}

We propose a ``minimal" fractional topological insulator (mFTI), motivated by the recent experimental report on the fractional quantum spin Hall effect in 
a transition metal dichalcogenide moir\'{e} system. The observed effect suggests the possibility of a topological state living in a pair of half-filled conjugate Chern bands with Chern numbers $C=\pm 1$.
We propose the mFTI as a novel candidate topological state in the half-filled conjugate Chern bands. The mFTI is characterized by the following features: (1) It is a fully gapped topological order (TO) with 16 Abelian anyons if the electron is considered trivial (32 including electrons);
(2) the minimally-charged anyon carries electric charge $e^\ast = e/2$, together with the fractional quantum spin-Hall conductance, implying the robustness of the mFTI's gapless edge state whenever time-reversal symmetry and charge conservation are present;
(3) the mFTI is ``minimal" in the sense that it has the smallest total quantum dimension (a metric for the TO's complexity) within all the TOs that can potentially be realized at the same electron filling and with the same Hall transports; the mFTI is also the {\it unique} minimal TO that respects time-reversal symmetry. (4) the mFTI is the common descendant of multiple valley-decoupled ``product TOs" with larger quantum dimensions. It can also be viewed as the result of gauging multiple symmetry-protected topological states. 
Similar mFTIs are classified and constructed for a pair of $1/q$-filled conjugate Chern bands. We further classify the mFTIs via the stability of the gapless interfaces between them.

\end{abstract}

\date{\today}

\maketitle

\section{Introduction}

The discovery of fractional Chern insulator~\cite{youngFCI,harvardFCI} and fractional quantum anomalous Hall insulator at zero magnetic field~\cite{xuFQAH1,xuFQAH2,makFCI,shFQAH,juFQAH} have led to a new excitement on strongly correlated states of matter in the moir\'{e} systems. 
Recent breakthrough in twisted bilayer MoTe$_2$ has found evidence of the fractional quantum spin Hall effect~\cite{FTIex}, motivating the consideration of a potential fractional topological insulator (FTI), a new form of topological quantum matter which could be fundamentally different from the previously known fractional quantum Hall states. The electronic structure that underlies the fractional quantum spin Hall effect consists of conjugate Chern bands, namely time-reversed pairs of Chern bands with opposite Chern numbers.

Conjugate Chern bands are common in moir\'{e} systems based on transition metal dichalcogenide (TMD) homobilayers with spin-valley locking\cite{wu2019topological,devakul2021magic,Crepel2024}. A typical scenario contains a Chern band with Chern number $C=+1$ at valley-1 (spin-up) and its time-reversed partner, a Chern band with $C=-1$, at valley-2 (spin-down). In the twisted bilayer MoTe$_2$ system experimentally studied in Ref. \onlinecite{FTIex}, single and double integer quantum spin Hall insulators were observed at total fillings $\nu_{\rm tot} = 2$ and $4$.
These insulators naturally correspond to one and two pairs of fully filled conjugate Chern bands with $C=\pm 1$, consistent with the band structure calculations\cite{WuBAB2024_mote2,Xiao2024_mote2,ZhangYang2024_mote2}. Note that the two pairs of conjugate Chern bands involved at $\nu_{\rm tot} = 4$ share the same chirality within each valley. The fractional quantum spin Hall effect was discovered at $\nu_{\rm tot} = 3$ where the first pair of conjugate Chern bands are fully filled and the second pair are half-filled. The observation includes nonlocal transport signals, the quantization of the edge conductance at fractional filling, and the suppression of the edge conductance when the time-reversal symmetry is explicitly broken by an in-plane magnetic field, analogous to the integer quantum spin Hall insulator~\cite{QSHex}. These observed effects (modulo the contribution from the fully filled bands) motivate the consideration of a novel time-reversal invariant FTI with fractionalized helical edge states and spin-Hall conductance $\sigma^{\rm sh} =1/2$ in a pair half-filled conjugate Chern bands. 
In addition to fractionalization, the FTI is further embellished by the rich symmetries of relevant moir\'{e} systems, including charge conservation, spin-$S_z$ conservation, and time-reversal symmetry\cite{FTIex}.

The current experimental results in twisted bilayer MoTe$_2$ are not yet sufficient to pin down the nature of the potential topological state. Also, conjugate Chern bands are ubiquitous in other moir\'{e} TMD systems as well. Hence, it is valuable to take a broader perspective and explore the landscape of possible FTI candidates in half-filled conjugate Chern bands.  
A possibility is that the FTI is simply a product of a pair of time-reversal conjugate topological orders (TO) in each valley, like some of the FTIs considered theoretically in the past~\cite{lidnerstern,cheng2012} (For a review of previous theoretical discussion of FTI, please refer to Ref.~\onlinecite{Stern_2016}. We also refer to the seminal works Ref. \onlinecite{FTI} and \onlinecite{FTI2} for general properties of generic FTIs). In the context of the TMD moir\'{e} system, the Chern band at valley-1 (valley-2) with Chern number $C = 1$ ($C=-1$) can be viewed as a Landau level due to the real space pseudo magnetic field~\cite{macdonaldLL,Paul_2023} (as illustrated in the upper panel of Fig. \ref{flux}). For a half-filled Landau level, the candidate topological order of electrons could be the non-Abelian Pfaffian state~\cite{mooreread,pfaffianwen,pfaffianwen2}, anti-Pfaffian state~\cite{Levin_AntiPf}, PH-Pfaffian state~\cite{Son2015},
the Abelian $\U(1)_8$ state~\cite{GREITER1992567}, or the ``$331$" state \cite{WenZee1992,OverboschWen2008}. Some of such fractional quantum Hall states and the closely related composite Fermi liquids have been numerically investigated as the candidate states for the valley-polarized half-filled Chern bands in twisted bilayer MoTe$_2$\cite{ZhangYang2024_mote2,ChoGilYoung_mote2_nonAbelian,Xiao2024_mote2,Fu2024NonAbelianTopoMini,Sheng2024nonAbelian,Dong2023CFL,Goldman2023CFL}. 
Hence, the potential (valley-unpolarized) FTI in a pair of half-filled Chern bands with $C=\pm1$ could be the product of one of the TOs mentioned above in one valley, and its time-reversal conjugate in the other valley. In such product TOs, the TOs from the two valleys essentially decouple. Understanding novel FTI candidates beyond such simple product TOs is both timely and important due to the potential relevance to the twisted bilayer MoTe$_2$ and the ubiquity of conjugate Chern bands in moir\'{e} TMD systems. Also, the inter-valley interaction in the moir\'e TMD systems is expected to be strong, which often leads to valley polarization (which is necessary for the integer and fractional quantum anomalous Hall effects). In the system where the fractional quantum spin Hall effect was observed, although we do not expect valley polarization, the inter-valley interaction could still drive the system into novel FTIs fundamentally different from the product TOs of the two valleys, which would be the states when the intervalley interaction is ignored.

In this work, we explore what kind of FTI in a pair of half-filled conjugate Chern bands with $C=\pm 1$ is the ``minimal" one, in the sense that it gives the desired experimental signals (especially the spin-Hall conductance) and also has the minimal total quantum dimension (which is, in simple terms, a metric for the topological order's complexity).
We take charge conservation, spin-$S_z$ conservation, and time-reversal symmetry into account in the search for the minimal FTI. It turns out that the minimal FTI (mFTI) is {\it unique}, and it is {\it NOT} a simple product of a pair of conjugate TOs of electrons from the two valleys, though it can be viewed as the common descendant of multiple ``product TOs" via anyon condensations. 
Our mFTI also has a spin-Hall conductance $\sigma^{\mathrm{sh}} = 1/2$, which is consistent with the experimental observation once the fully filled bands are included. This spin-Hall conductance is also half of the value for the elementary non-interacting quantum spin-Hall insulators~\cite{kaneqsh,zhangqsh}. The minimally charged anyon of the mFTI carries an electric charge $e^\ast = e/2$. In contrast, all the product TOs with the same response must have a minimal charge equal to or smaller than $e/4$. Additionally, according to the criterion established in Ref.~\onlinecite{FTI}, our mFTI has robust gapless edge states that remain protected by time-reversal symmetry and charge conservation even when $S_z$ is not conserved. 

If we only impose charge conservation, $S_z$ conservation, and the Hall transport signals (but not time-reversal symmetry) in the search, we find that the mFTI belongs to a list of minimal TOs following an 8-fold classification. 4 of the 8 minimal TOs, albeit time-reversal broken, are interesting non-Abelian TOs that exhibit quantized thermal Hall effects in addition to the required Hall responses associated with the charge and $S_z$ quantum numbers.

We generalize the construction of mFTI (and the minimal TOs) to a pair of conjugate $1/q$-filled Chern bands (for positive integer $q$). Under the symmetry and Hall transport constraints, mFTIs are found to be unique for a general $q$. We classify all mFTIs via the stability of the gapless interfaces between them.

\section{Construction of the mFTI}
\label{sec:mFTI_construction}

\begin{center}
\begin{figure}
\includegraphics[width=0.45\textwidth]{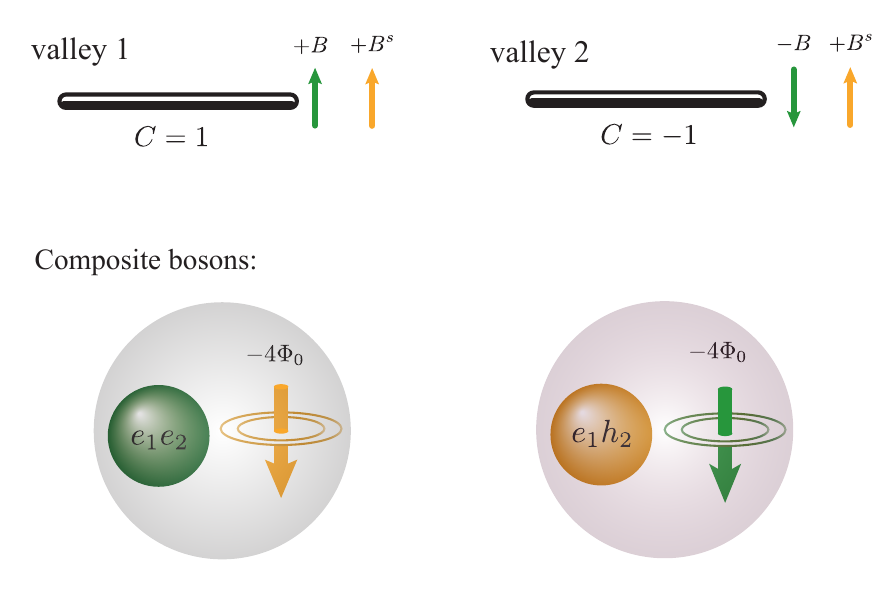}
\caption{Mutual flux attachment construction of the mFTI. Upper panel: The pair of conjugate Chern bands with Chern number $C=\pm 1$ are both half-filled. The green and yellow arrows indicate their contribution to the effective magnetic field seen by the inter-valley Cooper pair and the effective spin-magnetic field seen by the inter-valley excitons. In total, the Cooper pair sees no net magnetic field while the exciton sees a spin flux $\Phi^s = 2\Phi_0$ per moire unit cell. Lower panel: The inter-valley Cooper pair (green ball) is attached with flux $- 4\Phi_0$ seen by the spin-1 inter-valley exciton (yellow ball), forming a composite boson. Hence, the exciton sees a total zero magnetic field when each valley has a filling factor $\nu = 1/2$. The Cooper pair also sees the exciton as a $- 4\Phi_0$ flux. This flux attachment turns the exciton into another composite boson. $e_{1,2}$ and $h_{1,2}$ stand for the electrons and holes in the two valleys.} \label{flux}
\end{figure}
\end{center}

The mFTI may be perceived and constructed in various ways. Let us first explore the most intuitive flux-attachment picture of the mFTI, which is based on the kinematics of both the inter-valley Cooper pair, and also the inter-valley exciton. In the TMD system, the motion of electrons in each valley is governed by a real-space pseudo magnetic field. Hence, the physics of a Chern band with Chern number $C = +1$ can be viewed as a Landau level in a magnetic field~\cite{macdonaldLL,Paul_2023}. More precisely, an electron in valley-1(2) will see magnetic flux $\pm \Phi_0 = 2\pi$ in each moir\'{e} unit cell. If one performs the standard flux attachment to electrons (independently for each valley), it is most natural to construct the conjugate composite Fermi liquid state discussed recently~\cite{zhangccfl,nayanccfl,shi2024excitonic}, which is a compressible state in the bulk. However, there is another alternative flux attachment which naturally leads to an incompressible FTI. 

Let's start by considering the bosonic objects of the system: the inter-valley Cooper pairs and inter-valley excitons. A spin singlet inter-valley Cooper pair (whose annihilation operator is labeled as $b^c$) carrying charge-$2e$ sees zero net magnetic fields from the two valleys. An inter-valley exciton (labeled as $b^s$) carrying zero charges and spin-1 sees ``spin" flux $\Phi^s = 2\Phi_0$ in each moir\'{e} unit cell. $b^s$ should be viewed as a bosonic rotor. It can have both positive and negative density fluctuation but has zero average density when the system is time-reversal invariant. At half-filling of the conjugate pair of Chern bands, the density of $b^c$ is at $1/2$ particle per moir\'{e} unit cell. It is energetically favorable for each $b^c$ to be attached with flux $- 4\Phi_0$ seen by $b^s$, such that $b^s$ sees zero total spin flux. Likewise, $b^c$ will view each $b^s$ as $- 4\Phi_0$ flux. Here, we note that since $b^s$ has zero average density, $b^c$ still sees zero total flux. A schematic illustration of the flux attachment is shown in Fig. \ref{flux}. After the mutual flux attachment, the ``composite bosons", formed by $b^c$ and $b^s$ bound with fluxes, can condense. The condensate of the composite bosons is an incompressible state without any spontaneous symmetry breaking due to the attached gauge fluxes. 

The state constructed using the mutual flux attachment picture described above is described by the following Chern-Simons theory: 
\beqn \cL_{\mathrm{CS}} = \frac{4\ii}{2\pi} a^s d a^c + \frac{\ii}{2\pi} A^s d a^s + \frac{2 \ii}{2\pi} A^e d a^c. \label{CS}\eeqn Here $a^s d a^c$ is a short-handed notation for $\epsilon_{\mu\nu\rho}a^s_\mu \partial_\nu a^c_\rho$. $a^s_\mu$ and $a^c_\mu$ are the ``dual gauge field" of the current of $b^s$ and $b^c$ respectively: $J^{c,s}_\mu = \frac{1}{2\pi} \epsilon_{\mu\nu\rho} \partial_\nu a^{c,s}_\rho$. $A^e_\mu$ and $A^s_{\mu}$ are the background charge and ``spin gauge fields": $A^e_\mu = (A_{1, \mu} + A_{2,\mu})/2$, $A^s_\mu = A_{1,\mu} - A_{2,\mu}$. Here $A_{1,2}$ are introduced as the background gauge fields of valley-1 (spin-up) and valley-2 (spin-down) electrons. They arise from the conservation of electron number in each valley/spin species. If we integrate out the dynamical gauge fields $a^c$ and $a^s$ from Eq.~\eqref{CS}, the response theory for this state is \beqn \cL_{\mathrm{res}} = \frac{1}{2} \frac{\ii}{2\pi} A^e d A^s = \frac{1}{2} \frac{\ii}{4\pi} A_1 d A_1 - \frac{1}{2} \frac{\ii}{4\pi} A_2 d A_2, \label{res}\eeqn 
indicating the desired spin-Hall conductance $\sigma^{\mathrm{sh}} = 1/2$. Also, there are no ``diagonal" Hall responses for the electric charge and $S_z$ quantum numbers separately, which is consistent with time-reversal symmetry.

The mutual Chern-Simons theory Eq.~\eqref{CS} implies that the mFTI is actually a $\mathbb{Z}_4$ TO ``enriched" by two $\U(1)$ symmetries and time-reversal symmetry ${\cal T}$, i.e. it is a type of symmetry-enriched topological orders. As we will see below, the anyon data of this $\mathbb{Z}_4$ TO is equivalent to that of the $\mathbb{Z}_4$ toric code\cite{Kitaev2003} or a $\mathbb{Z}_4$ gauge field in 2+1D. In our context, the novel aspects of the $\mathbb{Z}_4$ TO lie in its symmetry enrichment and the fact that it is the unique minimal FTI compatible with the symmetry and transport requirements.
Throughout this work, we assume that ${\cal T}^2=-1$ when acting on an electron or a hole, i.e. electrons/holes are Kramers doublets. Within the $\mathbb{Z}_4$ TO, the ``minimally-charged" anyon $\mathbf{e}$ carries physical electric charge $e/2$, and the ``minimal-spin" anyon $\mathbf{m}$ carries a physical spin quantum number $S_z = \hbar/4$. Both of these anyons are self-bosons, but they see each other as a $\Phi_0/4$ flux, i.e. they have mutual statistical angle $\pi/2$. The $\mathbf{e}\mathbf{m}$ bound state carries an electric charge and a spin quantum number $(e/2, \hbar/4)$, and it is an anyon with self-statistical angle $\pi/2$, i.e. its topological spin is $\theta_{\bf em} = e^{\ii \pi/2} = \ii$. This $\mathbb{Z}_4$ TO also preserves the time-reversal symmetry ${\cal T}$, which keeps the anyon $\mathbf{e}$ invariant and maps the anyon $\mathbf{m}$ to its anti-particle $\mathbf{m}^3$. Curiously, $\mathbf{m}^2$ is a Kramers doublet under time reversal with ${\cal T}^2=-1$. This result can be seen using both the edge theory in Sec. \ref{sec:experiment} and the general arguments of Ref. \onlinecite{FTI}, which we will elaborate more on in Sec. \ref{bootstrap}.   

This $\mathbb{Z}_4$ TO have $16$ anyons modulo the electrons (and holes), which can be represented as composite states $\mathbf{e}^p\mathbf{m}^q$ 
with $p,q = 0, \cdots 3$. However, since this $\mathbb{Z}_4$ TO is constructed with bosonic objects $b^c$ and $b^s$, the background gapped electrons and holes were not yet taken into account. Hence, there exist another 16 anyons, which are bound states of $\mathbf{e}^p\mathbf{m}^q$ and the electron. Note that the bound state of $\mathbf{e}^p\mathbf{m}^q$ and the electron has a different topological spin from $\mathbf{e}^p \mathbf{m}^q$. Here, we've effectively included the electron as a ``transparent" Abelian anyon (that braids trivially with all other anyons), which is a convenient technical choice for showing the minimality of this $\mathbb{Z}_4$ TO later. With the electrons and holes included, the $\mathbb{Z}_4$ topological order has in total 32 Abelian anyons, resulting in a squared total quantum dimension $\cD^2 = 32$. In Sec.~\ref{bootstrap}, we will prove that $\cD^2 = 32$ is the minimal squared total quantum dimension compatible with the given Hall transports associated with the electric charge and spin quantum number $S_z$. If one views electrons/holes as trivial anyons and identifies the anyons differed only by an electron/hole, then the mFTI effectively has a squared total quantum dimension $\tilde{\cD}^2 = 16$. In the Appendix, we further prove that the mFTI proposed here is the {\it unique} mFTI compatible with charge conservation, $S_z$ conservation, time-reversal symmetry, and the desired Hall responses. We remark that our proof also shows that the mFTI remains to be the unique minimal TO even if we replace ${\cal T}$ by a ``non-Kramers" time-reversal symmetry ${\cal T}'$ (with ${\cal T}'^2 = 1$). However, the stability of the edge state of the mFTI is different under the two types of time-reversal symmetries, as discussed in the next section.

\section{Experimental signatures}
\label{sec:experiment}

Following the standard derivation of the edge states from the bulk CS theory~\cite{wenedge1,wenedge2}, we obtain the Lagrangian for the $1d$ edge based on the bulk CS theory Eq.~\ref{CS} \beqn \cL_{\mathrm{edge}} = \frac{4}{4\pi} \ii \partial_\tau \phi^c \partial_x \phi^s  + \frac{4}{4\pi} \ii \partial_\tau \phi^s \partial_x \phi^c + \cdots \label{edge1}, \eeqn
where the ``$\cdots$" is the kinetic energy that will be written in a different basis below. $\partial_x \phi^s$ and $\partial_x \phi^c$ are respectively the density of the Cooper pair $b^c$ and spin-1 exciton $b^s$ at the boundary. We have the identification $e^{\ii 4 \phi^c} \sim b^c$ and $e^{\ii 4 \phi^s} \sim b^s$. Under the time-reversal symmetry ${\cal T}$, the Cooper pair $b^c$ we introduced is invariant, while the spin-1 exciton operator $b^s$ shares the same behavior as the lowering operator of a spin-1/2 object, i.e. ${\cal T}b^s {\cal T}^{-1} = -b^{s\dag}$. Therefore, the time-reversal action on the edge theory is given by ${\cal T}: \phi^c \rightarrow -\phi^c, \phi^s \rightarrow \phi^s +\pi/4$.\footnote{In principle, one can consider a more general time-reversal action with $\phi^c \rightarrow -\phi^c + \alpha_0$ for some finite constant $\alpha_0$. However, this constant $\alpha_0$ can be absorbed by redefining the time-reversal symmetry as the original one followed by an extra charge U(1) rotation. The redefined time-reversal action remains to be a symmetry action of order 2. We caution that one cannot redefine the time-reversal symmetry by combining it with arbitrary spin $S_z$ rotations because the resulting action is generically not an order-2 action anymore.} The anyon ${\bf m}^2$ is identified with the field $e^{\ii 2\phi_s}$, which implies that ${\bf m}^2$ is a Kramers doublet under time reversal with ${\cal T}^2=-1$.

If the spin $S_z$ conservation is broken but time-reversal symmetry ${\cal T}$ is still preserved, an extra term $\cos(8 \phi_s)$ is allowed in the edge theory Eq.~\ref{edge1}. When this term is relevant, it will lead to two-fold degenerate ground states at the edge, which are characterized by nonzero expectation values of gauge invariant and time-reversal odd operator $\cos(4\phi_s)$, or $\sin(4\phi_s)$. Hence, the edge theory remains gapless unless there is either spontaneous or explicit time-reversal symmetry breaking (assuming charge is conserved). A different way to understand the stability of the gapless edge using only the bulk data will be given below. We comment that the ${\cal T}^2=-1$ is important for the stability of the gapless edge (in the absence of the $S_z$ conservation). A non-Kramers time-reversal symmetry $\cal T'$ would allow terms like $\cos(4\phi_s)$ or $\sin(4\phi_s)$ to gap out the edge without breaking $\cal T'$. In the following analysis, the time-reversal symmetry always refers to the Kramers-type time-reversal symmetry with ${\cal T}^2=-1$ unless specified otherwise.

By recombining $\phi^s$ and $\phi^c$ into $\phi_{1,2} = \phi^s \pm \phi^c$, we obtain a pair of counter-propagating (or helical) modes \beqn \cL_{\mathrm{edge}} = \frac{2}{4\pi} \ii \partial_\tau \phi_1 \partial_x \phi_1 - \frac{2}{4\pi} \ii \partial_\tau \phi_2 \partial_x \phi_2  + V_{IJ} \partial_x \phi_I \partial_x \phi_J. \label{edge2} \eeqn The density 
$\partial_x \phi_1$ (or $\partial_x \phi_2$) 
carries charge $e$ (or $-e$) and spin $\hbar/2$.  We have added the kinetic energy with the velocity matrix $V_{IJ}$ in Eq.~\eqref{edge2} which arises from the non-topological part of the system. If $V_{IJ}$ is proportional to %Pauli matrix $\sigma^z$
the identity matrix, the left and right counter-propagating modes do not interact with each other, and this edge theory would lead to a fractionally quantized $1d$ conductance per edge: \beqn G = \frac{1}{2}\frac{e^2}{h}.\eeqn 
In a two-terminal measurement, the two-terminal conductance receives contributions from two edges connecting the two leads. At the total filling factor $\nu_{\rm tot} =3$, there are two completely filled bands with $C = \pm 1$ in addition to the mFTI state in the half-filled conjugate Chern bands. Therefore, the total two-terminal conductance should be $G = 3 e^2/h$, which was observed experimentally in the twisted bilayer MoTe$_2$ at $\nu_{\rm tot} =3$.~\cite{FTIex}

To give a full analysis of the $1d$ edge conductance with counter-propagating fractionalized modes, one needs careful analysis of the physics in the $1d$ channel as well as in the metallic lead and the contact~\cite{kanefisherpol,kanefisher1,kanefisher2}. We expect the full analysis to be more involved than the edge state of the quantum spin-Hall insulator without fractionalization~\cite{xuedge,wuedge}, and we defer the systematic study to the future. 

Our mFTI is essentially a bosonic symmetry-enriched TO in the sense that although the system is made of correlated electrons/holes, a single electron/hole is always gapped in the bulk and the boundary. Hence, there is always a single particle gap in the system, despite the existence of the gapless charge modes at the edge, contrary to most fractional quantum Hall states. This effect is analogous to the bosonic SPT state proposed to be realized in the bilayer graphene~\cite{xubsptgraphene}.

The smallest electric charge carried by the anyons of the mFTI is $e^\ast=e/2$. The odd value of $\sigma^{\rm sh}/e^* = 1$ in the mFTI guarantees the gaplessness of the edge whenever time-reversal symmetry and charge conservation are present (even without the $S_z$ conservation).\cite{FTI} In contrast, using similar arguments as Ref. \onlinecite{FTI}, one can show that, given the spin-Hall conductance $\sigma^{\rm sh} =1/2$ (and the vanishing ``diagonal" Hall responses associated with the charge and spin quantum numbers), ANY TO that is a product of two decoupled TOs from the two valleys must have excitations with fractional electric charges equal to (or smaller) than $e/4$. As a concrete example, the product of a Pfaffian state in one valley and its time-reversed partner in the other valley admits anyons with $e/4$ electric charge~\cite{mooreread,pfaffianwen,pfaffianwen2}.

\begin{figure}
    \centering
    \includegraphics[width=0.9\linewidth]{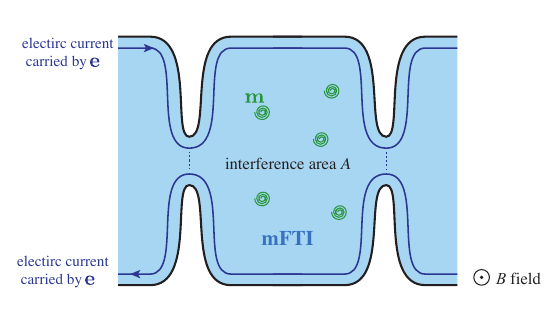}
    \caption{A schematic illustration of a Fabry-P\'erot interferometer with the mFTI. The upper and lower edges of the device are close to each other at two quantum point contacts, allowing the ${\bf e}$ to tunnel between them.} 
    \label{fig:FP}
\end{figure}

Based on the discussion above, we propose the following experiments that can distinguish the mFTI from other candidates, especially the product of decoupled topological states from the two valleys:
\begin{itemize}[leftmargin=*]
    \item Measurement of smallest charge excitations:
    
    The smallest fractional charge in topological states can be measured using multiple methods including the shot noise at quantum point contacts\cite{Saminadayar_ShotNoise,Saminadayar_ShotNoise}, local capacitance probe of localized charge states \cite{Yacoby_FractionalCharge_SET}, and Coulomb oscillations in antidots \cite{Kou_Antidots}. The smallest fractional charge in mFTI is $e/2$, while all the candidates formed by the products of valley-decoupled topological states must have charge-$e/4$ excitations. Therefore, the measurement of the smallest charge can distinguish the mFTI from other product topological states.
    
    \item  Anyon interferometry:

    Another characteristic feature of the mFTI is the braiding statistics $\theta_{\bf em} = e^{\ii \pi/2}$ between the anyons ${\bf e}$ and ${\bf m}$. These braiding statistics, as well as the smallest electric charge, can be measured using anyon interferometry devices\cite{KevilsonInterfero,WenHallInterfero,HalperinInterfero,NakamuraNatPhys2019,NakamuraNatPhys2020,NakamuraPRX2023,West2023PRX,West2009PRL,Camino2005PRB,Camino2007PRL,Mahalu2010PNAS,Young2024}. In our mFTI, the ${\bf e}$ and ${\bf m}$ anyons carry electric charge and spin respectively. For a Fabry-P\'erot interferometer with the mFTI, upper and lower edge currents surrounding the device interact with each other at two quantum point contacts (see Fig. \ref{fig:FP} for a schematic illustration). The edge currents are carried by the ${\bf e}$ anyons, and the conductance of the device will receive correction from both the Aharonov–Bohm effect, as well as the braiding statistics between the ${\bf e}$ anyons in the edge currents and the localized ${\bf m}$ anyons in the device.  The interference phase $\varphi$ of the current is given by \beqn \varphi = 2\pi \frac{e^*}{e} \frac{A B}{\Phi_0} + \frac{\pi}{2}  N_{\bf m} , \eeqn where $e^*=e/2$ is smallest charge in the mFTI. The first term corresponds to the Aharonov–Bohm effect proportional to the total external magnetic flux through the interferometer, where
    $A$ is the area of the interferometer, $B$ is the external out-of-plane magnetic field, 
    and $\Phi_0$ is the magnetic flux quantum. The second term arises from the braiding statistics between ${\bf e}$ anyons in the edge current, and localized ${\bf m}$ anyons in the interferometer. $N_{\bf m}$ is the number of ${\bf m}$ anyons in the interferometer region. The factor $\pi/2$ comes from the braiding statistics $\theta_{\bf em}$ between ${\bf m}$ and the smallest charge-carrying anyon ${\bf e}$. The conductance of the interferometer, which is periodic in $\varphi$, can be measured as a continuous function of the area $A$ and the magnetic field $B$. One can extract $e^*$ from the periodicity of the conductance as a function of $AB$. $N_{\bf m}$ can change discretely through anyon tunneling events, resulting in jumps in the phase $\varphi$, as the anyon ${\bf m}$ jumps in or out of the interferometer region.

    In contrast, for all the product TO candidates, the smallest charge $e^*=e/4$ takes a different value from the mFTI, resulting in a different periodicity of $\theta$ as a function of $AB$. The term $N_{\bf m} \frac{\pi}{2}$ also needs to be modified according to the candidate states. For example, for the product of $\U(1)_8$ and its time-reversed partner, this term would be $\frac{\pi}{4}$ times the number of the smallest-charge anyon. For the product of the Pfaffian state and its time-reversal partner, one expects an even-odd effect caused by the fact that, as far as the interference pattern is concerned, an even number of the non-Abelian Ising anyons collectively yield the same statistical phase as an Ableian anyon, while an odd number of them produce interference associated with the non-Abelian statistics \cite{Bonderson2008Interferometry,West2023PRX}.
\end{itemize}

% Therefore, the mFTI and the product TOs from the two valleys can be experimentally distinguished by probing the electric charge fractionalization. Examples of such probes include the shot noise of quantum point contacts\cite{Saminadayar_ShotNoise,Saminadayar_ShotNoise}, local capacitance probe of localized charge states \cite{Yacoby_FractionalCharge_SET}, and Coulomb oscillations in antidots \cite{Kou_Antidots}. Another key feature of the proposed mFTI is the braiding statistics $\theta_{\bf em} = e^{\ii \pi/2}$ between the anyons ${\bf e}$ and ${\bf m}$. These braiding statistics can be measured using anyon interferometry devices\cite{KevilsonInterfero,WenHallInterfero,HalperinInterfero}. 

\section{Bootstrapping the minimal FTI}

\label{bootstrap}

In this section, we bootstrap the minimal topological order (TO) allowed by the Hall responses under the charge and the $S_z$-spin U$(1)$ symmetries and show that $\cD^2 = 32$ is the minimal squared total quantum dimension compatible with a spin-Hall conductance $\sigma^{\rm sh}= 1/2$, zero electric Hall responses, and zero Hall response with respect to the spin $S_z$ quantum number. The vanishing of the diagonal Hall responses associated with either the electric charge or $S_z$ is required by the time-reversal symmetry. As we will see, the method used in this section is fundamentally based on the internal consistency of topological orders enriched by symmetries, and it yields the structure of the minimal TOs without relying on any microscopic details of the system. Hence, this method is a bootstrap-type approach. After we bootstrap the minimal TOs, we will discuss the physical significance of the minimal total quantum dimension.

For technical convenience, we prefer to re-organize the charge and spin U(1) symmetries into $\U(1)_\uparrow\times$ U(1)$_\downarrow$, where $\U(1)_{\uparrow,\downarrow}$ are associated with the charge conservation within each of the valleys 1 and 2 (which are locked to electron spins $\uparrow$ and $\downarrow$). The Hall responses we focus on in this paper are equivalent to the combination of (1) $\U(1)_{\uparrow}$-Hall conductance $\sigma^{\rm h}_1=1/2$, (2) $\U(1)_{\downarrow}$-Hall conductance $\sigma^{\rm h}_2=-1/2$, and (3) zero mixed $\U(1)_{\uparrow} \times \U(1)_{\downarrow}$ response. For an anyon $x$, we denote its fractional charge under $\U(1)_{\uparrow} \times \U(1)_{\downarrow}$ as $q_x= (q_{x,1}, q_{x,2})$, which is defined modulo the charge of local bosonic objects, i.e. $q_x \sim q_x + (n,m)$ for $n,m\in \mathbb{Z}$ with $m+n$ even. The fractional electric charge and $S_z$-spin of the anyon $x$ is then given by $(q_{x,1}+ q_{x,2})e$ and $(q_{x,1} - q_{x,2})\hbar/2$. In the TO, the electrons and holes are all identified as the transparent fermionic particle $f$ with $q_f = (\pm 1,0)\sim (0,\pm 1)$. Two such $f$ particles fuse into a trivial anyon $\mathbf{1}$, i.e. $f\times f = {\bf 1}$ 

To bootstrap the minimal TO, we start by considering the anyons $v_{1,2}$ created by the adiabatic insertion of the $2\pi$ flux of $\U(1)_{\uparrow,\downarrow}$. They must be Abelian anyons\cite{Cheng_translation}. Their fractional charges are given by the Hall conductivities $\sigma^{\rm h}_{1,2}$: $q_{v_1}= (1/2,0)$ and $q_{v_2}= (0,-1/2)$. Their topological spins are $\theta_{v_1} = e^{\ii\pi \sigma^{\rm h}_1} = \ii$ and $\theta_{v_2} = e^{\ii\pi \sigma^{\rm h}_2}=-\ii$, which are the consequences of the Aharonov-Bohm (AB) phase between charge and flux. Based on the AB effect, the mutual statistics between $v_1$ and $v_2$ must be trivial because $v_1$ ($v_2$) does not carry any fractional $\U(1)_\downarrow$-charge ($\U(1)_\uparrow$-charge). The anyons $v_{1,2}^2$ are associated with the $4\pi$ flux of $\U(1)_{\uparrow,\downarrow}$. They are both self-bosons, i.e. $\theta_{v_1^2} = \theta_{v_1}^4 = 1$ and $\theta_{v_2^2} = \theta_{v_2}^4 = 1$. They have charges $q_{v_1^2} = (1,0)$ and $q_{v_2^2} = (0,-1)$, which is non-trivial compared to local bosonic objects. Hence, $v_1^2$ and $v_2^2$ are non-trivial anyons. Since $\theta_{v_1^2} =  \theta_{v_2^2}$ and $q_{v_1^2} \sim q_{v_2^2}$, there are two scenarios: (1) $v_1^2$ and $v_2^2$ belongs to the same anyon type, and (2) $v_1^2$ and $v_2^2$ are different anyons. 

For the second scenario with $v_1^2 \neq v_2^2$, there are at least 32 different anyons of the form $v_1^n v_2^m f^k$ with $n,m=0,1,2,3$ and $k=0,1$. They can be distinguished from each other based on their topological spins, fractional charges, and the assumption that $v_1^2 \neq v_2^2$. Moreover, based on the AB-phase, $v_2^2$ and $v_1^2$ braids trivially with any of the 32 anyons listed above. For a TO to be ``complete"\cite{kitaev_2006}, there must be some extra anyons that braid non-trivially $v_1^2$ and $v_2^2$. Consequently, the squared total quantum dimension must be larger than 32 in this scenario. Therefore, to search for the minimal TO, we only need to focus on the first scenario. 

For the first scenario with $v_1^2 = v_2^2$, we show below that the minimal squared total quantum dimension is exactly $\cD^2 = 32$. Let $s$ denote the minimal (positive) integer $s$ such that $v_1^{2s} = v_2^{2s}$ equals the trivial anyon ${\bf 1}$. Given that $v_1^2 = v_2^2$ is non-trivial, $s\geq2$. In the spirit of bootstrapping the minimal TO, we should focus on the case with $s=2$. Similar arguments as below, when applied to the cases with $s \geq 3$, will lead to a squared total quantum dimension larger than 32. 

From now on, we focus on the first scenario with $s=2$. A list of 8 distinct Abelian anyons purely consists of $2\pi\mathbb{Z}$ fluxes of $\U(1)_{\uparrow,\downarrow}$ is given by
\begin{align}
  V= \{ \mathbf{1}, v_1, v_1^2,v_1^3, v_2, v_2^3, v_1v_2, v_1^3v_2 \}. 
\end{align}
By including the transparent fermion $f$, one finds 16 different Abelian anyons $\{\mathbf{1} ,f\} \times V$. They can be distinguished by their topological spins and fractional charges. Note that $v_1^2$ braids trivially with all anyons in $\{\mathbf{1} ,f\} \times V$. Therefore, there must be an extra anyon, called it $y$, that braids non-trivially with $v_1^2 = v_2^2$. However, it must braid trivially with $v_1^4 = v_2^4 = \mathbf{1}$. Using the AB-phase interpretation of the braiding with $v_1^n$ and $v_2^m$, we conclude that the fractional charge of $y$ must be $q_y = (1/4 + \mathbb{Z}, 1/4 + \mathbb{Z})$. Without loss of generality, we can pick 
\begin{align}
    q_y = (1/4 , 1/4).
    \label{eq:qy}
\end{align}
For other choices of $q_y$, we can find another anyon with the fractional charge $(1/4, 1/4)$ by fusing $y$ with one of the Abelian anyons in $\{\mathbf{1} ,f \} \times V$. From the AB effect, we can obtain mutual statistics between $y$ and $v_1^n v_2^m f^k$:
\begin{align}
    M_{y, v_1^n v_2^m f^k} = e^{\ii \frac{2\pi}{4} (n+m)}.
\end{align}
Therefore, the topological spins of the anyons $y v_1^n v_2^m f^k$ are all related to the topological spin of $y$ via
\begin{align}
    \theta_{y v_1^n v_2^m f^k} = \theta_y \times (\ii)^{n^2} (-\ii)^{m^2}e^{\ii \frac{2\pi}{4} (n+m)} (-1)^k.
\end{align}
Using this result of the topological spins (and fractional charges), we can examine the set of anyons $\{1,y\}\times \{ 1,f \}\times V$. This set may have redundancy because $y$ and $y v_1^2 f$ share the same topological spin and fractional charge. As a result, there are two types, type-I and type-II, of minimal TOs. 

For {\it type-I} minimal TOs, $y$ and $y v_1^2 f$ are distinct. Consequently, there are 32 different anyons in 
\begin{align}
    {\cal C}_{\rm I} = \{1,y\}\times \{ 1,f \}\times V.
\end{align}
For minimality, ${\cal C}_{\rm I}$ should be the entire set of anyons, and each anyon should be Abelian, which yields a squared total quantum dimension $\cD^2 =32$. The $\mathbb{Z}_4$ FTI constructed in Sec. \ref{sec:mFTI_construction} provides an example of such type-I minimal TOs. $v_1$, $v_2$, and $y$ correspond to the anyons $\mathbf{e} \mathbf{m}$ and $\mathbf{e}^3 \mathbf{m}$, $\mathbf{e}$ respectively. Note that the $\mathbb{Z}_4$ FTI not only has the desired Hall responses but also preserves the time-reversal symmetry. Moreover, ${\bf m}^2$, which corresponds to $v_1 v_2$, is the anyon generated by the insertion of a $2\pi$ flux in each valley. Its time-reversal conjugate should be associated with a $-2\pi$ flux in each valley, which still belongs to the same anyon type ${\bf m}^2$ given that $v_1^2 v_2^2 \sim {\bf 1}$. Knowing that ${\bf m}^2$ is related to its time-reversal partner by a $4\pi$ flux in each valley and that the spin Hall conductance is $\sigma^{\rm sh} = 1/2$, ${\bf m}^2$ must be a Kramers doublet under time reversal \cite{FTI}, which is consistent with the analysis in Sec. \ref{sec:experiment}. This statement has used the fact that the system is ultimately built out of the electron and holes that are Kramers doublets with ${\cal T}^2 = -1$.

For {\it type-II} minimal TOs, $y$ and $y v_1^2 f$ are identical. In other words, the fusion of $y$ and $v_1^2 f$, a self-fermion, yields $y$. Consequently, $y$ is a non-Abelian anyon whose quantum dimension of $y$ is at least $\sqrt{2}$. Now, there are 24 distinct anyons (or 12 if one treats the electron/hole $f$ as a trivial anyon):
\begin{align}
    {\cal C}_{\rm II} = (\{ 1,f \}\times V) \cup (y\times V).
\end{align}
For minimality, ${\cal C}_{\rm II}$ should be the entire set of anyons, and the quantum dimension of each anyon in $y\times V$ should be exactly $\sqrt{2}$, which again yields a squared total quantum dimension of $\cD^2 = 32$. Note that all the anyons in $y\times V$ share the same quantum dimension because of the Abelian nature of anyons in $V$. Such type-II minimal TOs must be non-Abelian. They also require non-trivial couplings between the two valleys. More importantly, it turns out that they must break the time-reversal symmetry ${\cal T}$, as we showed in the Appendix (where a full 8-fold classification of all minimal TOs is also given). In the next section, we briefly discuss examples of such type-II minimal TO.

We emphasize that the bootstrap of type-I and type-II minimal TOs only uses charge conservation, spin-$S_z$ conservation, and the Hall transport signals. In the Appendix, we further show that once time-reversal symmetry ${\cal T}$ is considered, the $\mathbb{Z}_4$ FTI constructed in Sec. \ref{sec:mFTI_construction} is the {\it unique} mFTI with the minimal total quantum dimension. Moreover, the analysis above fixed the anyon statistics of type-I and type-II minimal TOs up to the topological spin $\theta_y$. The full classifications of all the minimal TOs and the allowed values of $\theta_y$ are provided in the Appendix.

Now, we comment on the physical implication of the minimal total quantum dimension. The squared total quantum dimension ${\cal D}^2$ of a TO is a metric of the complexity of the TO, given by the sum of the squared quantum dimensions of all the anyons in the TO. It is directly related to the topological entanglement entropy that characterizes the long-range entanglement in the two-dimensional TO.\cite{KitaevPreskillTEE,LevinWenTEE} Therefore, the search for the minimal TOs is equivalent to identifying the TOs with the least amount of long-range entanglement allowed by the symmetry and transport requirements. One may expect the minimal TOs to be, in general, more robust than the more complex ones that require more long-range entanglement. Additionally, for Abelian TOs, ${\cal D}^2$ is equal to the number of anyons and, consequently, the ground state degeneracy on the torus (when the electron is treated as a trivial anyon). Therefore, generically speaking, if a minimal TO is Abelian, it must also be the TO with the smallest ground state degeneracy amongst all the Abelian TOs satisfying the symmetry and transport requirements. However, we caution that the total quantum dimensions of non-Abelian TOs are different from their ground state degeneracies on the torus. In the current context of minimal TOs with spin-Hall conductance $\sigma^{\rm sh} = 1/2$, we notice that the Abelian mFTI and other type-I minimal TOs have the same ground state degeneracy 16 on the torus, while the type-II minimal TOs' ground state degeneracies are 12. Despite the difference in ground state degeneracies, all the minimal TOs share the same total quantum dimensions.

As a side note, the previous results in fractional quantum states under a magnetic field provide further justification for the general idea of bootstrapping TOs with minimal quantum dimensions that are compatible with observed transport. The celebrated $1/m$ Laughlin states (with $m$ odd) are the TOs with the minimal quantum dimensions compatible with electrical Hall conductance $1/m$. For a half-filled Landau level, all of the leading incompressible fractional quantum Hall candidates, including the Pfaffian state~\cite{mooreread,pfaffianwen,pfaffianwen2}, anti-Pfaffian state~\cite{Levin_AntiPf}, PH-Pfaffian state~\cite{Son2015},
the Abelian $\U(1)_8$ state~\cite{GREITER1992567}, and the ``$331$" state \cite{WenZee1992,OverboschWen2008}, are the minimal TOs compatible with electrically Hall conductance $1/2$ (modulo integers). Therefore, bootstrapping the minimal TOs using transport and symmetry can be a powerful tool for the study of topological quantum matter.

\section{Alternative constructions of the mFTI}

\subsection{Descending from larger topological orders}
\label{sec:descend}

\begin{center}
\begin{figure}
\includegraphics[width=0.5\textwidth]{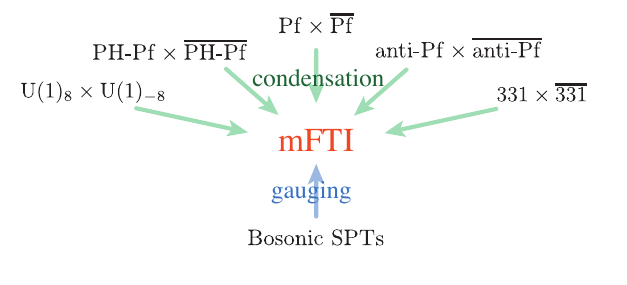}
\caption{The mFTI we propose is the common descendant of multiple product TOs with larger quantum dimensions. ``Pf" stands for Pfaffian. It can also be viewed as a symmetry-enriched TO
promoted from multiple bosonic SPTs by ``gauging" the discrete subgroup of the symmetries. } \label{descend}
\end{figure}
\end{center}

In this section, we present other constructions of the same $\mathbb{Z}_4$ mFTI. The mFTI can be constructed through a ``top-down" approach by descending from ``larger" TOs, such as a product of two separate TOs in each valley. The simplest example of such TO, which can be realized at the $\nu = 1/2$ filling of both valleys, is the $\U(1)_8 \times \U(1)_{-8}$ state. The $\U(1)_8$ TO realized in a half-filled Landau level can be viewed as a $\nu_b = 1/8$ bosonic Laughlin state for Cooper pairs~\cite{GREITER1992567}, described by a level-8 Chern-Simons theory for a dynamical U(1) gauge field $a$, whose flux is dual to the density of the {\it intra-valley} Cooper pair. The anyons of the $\U(1)_8 \times \U(1)_{-8}$ state correspond to the gauge charges $(n_1, n_2)$ of the two gauge fields, whose topological spins are given by $e^{\ii \pi(n_1^2 - n_2^2)/8}$. 
This anyon also carries electric charge $(n_1 - n_2)e/4$, and spin $S_z = (n_1 + n_2)\hbar/8$. 

The anyon $(4, 4)$ is obviously a self-boson, hence it can condense. The anyon $(4, 4)$ can be created by adiabatically inserting $4\pi$ fluxes of 
$\U(1)_\uparrow$ and $\U(1)_\downarrow$ in both valleys. It carries spin $S_z = \hbar$. To avoid spontaneously breaking the spin $S_z$ symmetry in the condensate, we bind the $(4, 4)$ anyon with the physical inter-valley exciton $b^s$ with $S_z = - \hbar$ to neutralize its $S_z$ spin. The mutual braiding statistics between the neutralized bosonic anyon $(4, 4)$ and a general anyon $(n_1, n_2)$ is 
$e^{\ii \pi (n_1 - n_2) }$. Hence, the condensate of $(4, 4)$ would confine all the anyons whose $n_1$ and $n_2$ have opposite parity. The condensate also identifies anyons $(n_1, n_2)$ and $(n_1 + 4, n_2 + 4)$. Consequently, there are, in total, 16 anyons left (before including the electrons/holes), the same as the mFTI. Using the notation of Sec. \ref{bootstrap}, the anyon $v_1$ and $v_2$ correspond to $(2, 0)$, and $(0, 2)$. One of the most elementary deconfined anyons is $(1, 1)$, which is a self-boson with electric charge $0$ and spin $S_z = \hbar/4$. Hence, $(1,1)$ is identified with the anyon $\bf{m}$ of the mFTI. Similarly, the anyon $(1, -1)$ can be identified as anyon $\bf{e}$ of the mFTI. The anyons $(1,1)$ and $(1, -1)$ also have the desired mutual braiding statistics $e^{\ii \pi/2}$. At this point, we can confirm that the condensate of the neutralized anyon $(4, 4)$ is the mFTI with $\mathbb{Z}_4$ TO.
% angle $\pi/2$. 
As an alternative approach, how the $\U(1)_8 \times \U(1)_{-8}$ TO descends, via the anyon condensation of (4,4), to the desired mFTI can also be shown explicitly using the $K$-matrix formalism.

In fact, one can show that if we start with the other obvious product TOs, such as the $\mathrm{Pfaffian} \times \overline{\mathrm{Pfaffian}}$, $331 \times \overline{331}$, 
$\text{anti-Pfaffian} \times \overline{\text{anti-Pfaffian}}$, or $\text{PH-Pfaffian} \times \overline{\text{PH-Pfaffian}}$, there is always a self-bosonic anyon created by inserting $4\pi$ fluxes in both valleys. Then, condensing this anyon after neutralizing its $S_z$ spin will always lead to the {\it same} mFTI state, as illustrated in Fig. \ref{descend}. Interestingly, a similar construction can be performed on the surface of a class-CII topological insulator where the $\mathbb{Z}_4$ TO appears as the anomalous surface TO. \cite{Potter2017}

As an aside, we can perform a similar ``descending" construction starting from a product TO with {\it different} Abelian TOs in the two valleys, such as $331\times \U(1)_{-8}$. A similar anyon condensation in  $331\times \U(1)_{-8}$ results in a type-I minimal TO that breaks time-reversal symmetry (manifested by its net chiral central charge $c=1$). Hence, the so-constructed type-I minimal TO is not a candidate for mFTI even though they share the same Hall responses with respect to the electric charge and the $S_z$ quantum number. All such type-I minimal TOs are classified in the Appendix.

\subsection{``Promoting" SPT states}

As we remarked earlier, the mFTI is a symmetry-enriched topological (SET) order. The subject of SET has attracted enormous theoretical interest in the past decade~\cite{hermeleSET,ranSET,wenSET,luSET}. 
One very general way to construct nontrivial SET states is by ``gauging" a part of the discrete symmetries of symmetry-protected topological (SPT) states~\cite{wenspt1,wenspt2}, i.e. by ``promoting" a part of the discrete global symmetries to local gauge invariance, and coupling the state to the gauge fields accordingly. 

In our current case, the mFTI can be constructed through the same procedure, as indicated in Fig. \ref{descend}. The SPT state we start with could be constructed by the inter-valley exciton operator $b^s$, and ``one quarter" of the inter-valley Cooper pair $b^c$. In other words, we formally fractionalize $b^c$ into four bosonic partons each carrying charge-$e/2$: $b^c = (\tilde{b}^c)^4$. Then, $\tilde{b}^c$ must coupled to a $Z_4$ gauge field. Now, we construct a minimal level-1 bosonic SPT state between $b^s$ and $\tilde{b}_c$~\cite{levinsenthil}, which can be described in various formulations, including the Chern-Simons theory~\cite{luashvin}, or the nonlinear sigma model~\cite{levinsenthil,xusenthil}. The most convenient formulation for our current purpose is the Chern-Simons theory developed in Ref.~\onlinecite{luashvin}. The Chern-Simons theory for many $(2+1)d$ bosonic SPT states takes a universal form: \beqn \cL_{\mathrm{spt},1} = \frac{\ii}{2\pi} \tilde{a}^c d a^s + \frac{\ii}{2\pi} A^s d a^s + \frac{1}{2}\frac{\ii}{2\pi} A^e d\tilde{a}^c, \label{spt1} \eeqn where $\tilde{a}^c_\mu$ and $a^s_{\mu}$ are the dual of the currents of $\tilde{b}^c$ and $b^s$. Now, we need to gauge this SPT by coupling $\tilde{b}^c$ to a $Z_4$ gauge field, which can be captured by adding the following terms \beqn \cL_{g,1} = \frac{\ii}{2\pi} c_1 d\tilde{a}^c   + \frac{4\ii}{2\pi} c_1 d c_2. \label{g1} \eeqn The gauge fields $c_{1,\mu}$ and $c_{2,\mu}$ are introduced as auxiliary dynamical gauge fields that describe the $Z_4$ gauge field through the last mutual Chern-Simons term. The coupling $\frac{\ii}{2\pi} c_1 d\tilde{a}^c$ is the minimal coupling between the $Z_4$ gauge field and the current of $\tilde{b}^c$. Combining the Lagrangians of Eq.~\eqref{spt1} and Eq.~\eqref{g1}, the Chern-Simons couplings of all the dynamical gauge fields $a_I = (\tilde{a}^c, a^s, c_1, c_2)$ can be organized into a 
% can be unified in a general form with a 
$K$-matrix theory \beqn \cL_{\mathrm{cs},1} = \frac{\ii}{4\pi} K^{IJ}_1 a_I d a_J, \ \ K_1 =
\begin{pmatrix}
0 & 1 & 1 & 0 \\
1 & 0 & 0 & 0 \\
1 & 0 & 0 & 4 \\
0 & 0 & 4 & 0
\end{pmatrix}.
\label{CS1} \eeqn

The charge-1 particle of the gauge field $c_1$ is the minimal $\mathbf{e}$ anyon of the $Z_4$ gauge field, it carries a $2\pi$ flux of $\tilde{a}^c$, which carries physical electric charge $e/2$, as expected. The charge-1 particle of gauge field $c_2$ carries $\pi/2$ flux of $c_1$, which is equivalent to $\pi/2$ flux of $a^s$ through the equation of motion of the gauge fields. The $\pi/2$ flux of $a^s$ carries $S_z$-spin $\hbar/4$, hence the charge-1 of $c_2$ corresponds to the anyon $\mathbf{m}$ of our mFTI.

To make a direct connection to the original CS theory Eq.~\eqref{CS}, we note that the $2\times 2$ block of the $K$-matrix associated with $(\tilde{a}^c, a^s)$ has determinant $-1$. Hence, $\tilde{a}^c$ and $a^s$ can be integrated out safely without compromising the nature of the TO. After integrating out $(\tilde{a}^c, a^s)$, and shifting $c_1 + \frac{1}{2} A^e \rightarrow c_1$, the theory Eq.~\ref{CS1} returns to the form of Eq.~\eqref{CS}, except now $(c_1, c_2)$ plays the role of $(a^s, a^c)$.

Another way of constructing the same mFTI state is by introducing the ``half-partons" for both $b^c$ and $b^s$: $b^c \sim (\tilde{b}^c)^2$, and $b^s \sim (\tilde{b}^s)^2$. Now, $\tilde{b}^c$ and $\tilde{b}^s$ carry electric charge $e$ and quantum number $S_z = \hbar/2$, respectively. And, they are both coupled to their own $Z_2$ gauge fields. We can make $\tilde{b}^c$ and $\tilde{b}^s$ form a level-1 bosonic SPT state. The entire theory now reads \beqn \cL_{\mathrm{cs},2} &=& \frac{\ii}{2\pi} \tilde{a}^c d \tilde{a}^s + \frac{\ii}{2\pi} (A^e + c_1) d \tilde{a}^c + \frac{2\ii}{2\pi} c_1 d c_2 \cr\cr &+& \frac{\ii}{2\pi} (\frac{1}{2}A^s + c'_1) d \tilde{a}^s + \frac{2\ii}{2\pi} c'_1 d c'_2. \eeqn Again, $\tilde{a}^c$ and $\tilde{a}^s$ are dual to the currents of $\tilde{b}^c$ and $\tilde{b}^s$ respectively, and they can be integrated out safely. After integrating out $\tilde{a}^c$ and $\tilde{a}^s$, the $\cL_{\mathrm{cs},2}$ reduces to a CS theory for $(c_1, c_2, c_1', c_2')$ with $K$-matrix \beqn K_2 = \begin{pmatrix}
0 & 2 & 1 & 0 \\
2 & 0 & 0 & 0 \\
1 & 0 & 0 & 2 \\
0 & 0 & 2 & 0
\end{pmatrix} \ \simeq \ K'_2 = \begin{pmatrix}
0 & 1 & 0 & 0 \\
1 & 0 & 0 & 0 \\
0 & 0 & 0 & 4 \\
0 & 0 & 4 & 0
\end{pmatrix}. \eeqn One can show that the $K_2$ matrix is equivalent to the $K'_2$-matrix up to an SL$(4, \mathbb{Z})$ transformation, and $K_2'$ again describes a $Z_4$ gauge field. The last two components of gauge fields of the $K'_2$-matrix couple to the external gauge field $A^e$ and $A^s$ in the same way as $a^c$ and $a^s$ in Eq.~\eqref{CS}.

\subsection{Example of type-II minimal TO}
\label{typeII}

The non-Abelian ``type-II minimal TO" discussed in Sec. \ref{bootstrap} can be obtained as descendants of the product TOs (with larger total quantum dimensions). As an example, We start from the product TO $\mathrm{Pfaffian} \times \U(1)_{-8}$. The anyons of the Pfaffian TO is a subset of the topological order $\mathrm{Ising} \times \U(1)_8$. Here, ``Ising" stands for the Ising TO, which has three anyons $\{\mathds{1},\sigma,\psi\}$ with $\mathds{1}$ the trivial anyon, $\sigma$ the Ising anyon with quantum dimension $\sqrt{2}$, and $\psi$ the self fermion. Their topological spins are $\theta_\mathds{1} = 1$, $\theta_\sigma = e^{\ii 2\pi/16}$, and $\theta_\psi = -1$ respectively. For the $\U(1)_8 \times \U(1)_{-8}$ sector of $\mathrm{Pfaffian} \times \U(1)_{-8}$, we can still use the charge $(n_1, n_2)$ of the gauge fields of the $\U(1)_8 \times \U(1)_{-8}$ Chern-Simons theory as a part of the anyon label. The contribution to the topological spins from the gauge charge $(n_1, n_2)$ is $\theta_{(n_1,n_2)} = e^{\ii 2\pi (n_1^2-n_2^2)/16}$. The full set of anyons of $\mathrm{Pfaffian} \times \U(1)_{-8}$ is given by
\begin{align}
    (\mathds{1},n_1, n_2) {\rm ~~~with~~~} n_1 =0,2,4,6 {\rm ~and~} n_2=0,1,...,7, \nonumber \\
    (\psi,n_1, n_2) {\rm ~~~with~~~} n_1 =0,2,4,6 {\rm ~and~} n_2=0,1,...,7, \nonumber \\
    (\sigma,n_1, n_2) {\rm ~~~with~~~} n_1 =1,3,5,7 {\rm ~and~} n_2=0,1,...,7,
\end{align}
where $(\psi, 4, 0)$ is identified as the transparent fermion $f$, i.e. the electron or the hole. The electric charge and $S_z$ quantum numbers of these anyons are given by $(n_1 - n_2) e/4$ and $(n_1+n_2)\hbar/8$. Note that $(\mathds{1}, 4, 4)$ is self-boson without electric charge. Its $S_z$ quantum number can be neutralized by adding a local boson, i.e. an intervalley exciton. 

By condensing the neutralized version of $(\mathds{1}, 4, 4)$ in $\mathrm{Pfaffian} \times \U(1)_{-8}$, we obtain a type-II minimal TO with 24 deconfined anyons. Using the notation of Sec. \ref{bootstrap}, the 16 Abelian anyons in $\{1,f\} \times V $ are given by
\begin{align}
  (\mathds{1},n_1, n_2) {\rm ~and~} (\psi,n_1, n_2) {\rm ~~~with~~~} n_{1,2} = 0,2,4,6
\end{align}
with the identification $(\mathds{1}/\psi,n_1, n_2) \sim (\mathds{1}/\psi,n_1+4, n_2+4)$.\footnote{$n_{1,2}$ should be viewed as integers mod $8$} In particular, we can identify $v_1 = (\mathds{1}, 2, 0)$ and $v_2 = (\mathds{1}, 0, 2)$. The $8$ non-Abelian quantum-dimension-$\sqrt{2}$ anyons are given by
\begin{align}
    (\sigma,n_1, n_2) {\rm ~~~with~~~} n_{1,2} =1,3,5,7
\end{align}
with the identification $(\sigma,n_1, n_2) \sim (\sigma,n_1+4, n_2+4)$. In particular, we have $y = (\sigma,1, 7) $.

This non-Abelian TO, described by the condensate of $(\mathds{1}, 4, 4)$, is one of the type-II minimal TO that can be realized in a pair of half-filled conjugate Chern bands. It has the same Hall responses with respect to the electric charge and the $S_z$ quantum number as the mFTI. However, this state breaks the time-reversal symmetry because it has the same chiral central charge $c = 1/2$ as the original $\mathrm{Pfaffian} \times \U(1)_{-8}$ TO. Consequently, this type-II minimal TO also has a quantized thermal-Hall response given by its chiral central charge $c= 1/2$.

More examples of type-II minimal TOs can be constructed as the descendants of other combinations of a non-Abelian TO (which could be the Pfaffian TO, the anti-Pfaffian TO, the PH-Pfaffian TO, and their conjugates) in one valley and an Abelian TO (which could be $\U(1)_8$, the 331 state, and their conjugates) in the other valley. A systematic construction of all the type-II minimal TOs is given in the Appendix. All four of them break time-reversal symmetry.

\subsection{mFTIs with $\sigma^{\rm sh}=\frac{1}{q}$: bootstrap and classification}
\label{sec:mFTI_q}

Now, we generalize the study of mFTI to a pair of conjugate Chern bands with $1/q$ electron filling in each band. Here, $q$ is a positive integer. It is natural to expect such a mFTI to have a fractional spin-Hall conductance $\sigma^{\rm sh} =1/q$ and vanishing diagonal Hall responses for the electric charge and the spin-$S_z$ quantum number. The symmetry of such a mFTI includes the charge conservation, $S_z$ conservation, and time-reversal symmetry ${\cal T}$. Per our definition, the mFTI is minimal in the sense that it has the smallest total quantum dimension among all the TOs with the required symmetries and Hall transport signals with the given $q$. We will separately discuss the mFTIs with even and odd $q$'s. And we will provide a classification of them.

For even $q$, by generalizing Sec. \ref{bootstrap}, we can show that there is a list of, in total, eight different minimal TOs compatible with charge conservation, $S_z$ conservation, and the Hall transports described above. The details are given in the Appendix. After further imposing time-reversal symmetry ${\cal T}$, the list narrows down to a {\it unique} mFTI, which has the $\mathbb{Z}_{2q}$ TO, the same TO as the $\mathbb{Z}_{2q}$ toric code. The squared total quantum dimension (with the electron included) is $\cD^2 = 8q^2$. The minimally-charged anyon has an electric charge $e^\ast=e/q$, and the minimal-$S_z$ anyon has the $S_z$ quantum number $\hbar/{2q}$. Such an mFTI is not a valley-decoupled product TO. Since $\sigma^{\rm sh}/e^\ast$ is odd, based on the result of Ref. \onlinecite{FTI}, the gapless edge state of the mFTI is protected by charge conservation and time-reversal symmetry (even when $S_z$ is not conserved). 
The edge theory is a straightforward generalization of Eq. \eqref{edge1}
\beqn \cL_{\mathrm{edge}} = \frac{2q}{4\pi} \ii \partial_\tau \phi^c \partial_x \phi^s  + \frac{2q}{4\pi} \ii \partial_\tau \phi^s \partial_x \phi^c + \cdots \label{edge1q}, \eeqn
Again, $e^{\ii 2q\phi^c}$ is the annihilation operator of the charge-$2e$ Cooper pair, while $e^{\ii 2q\phi^s}$ is that of an intervalley exciton with the $S_z$ quantum number $\hbar$. Under time-reversal symmetry, the edge modes transform as 
\begin{equation}
    \cT \phi_c \cT^{-1}=-\phi_c, \cT \phi_s \cT^{-1}=\phi_s+\frac{\pi}{2q}.
\end{equation}

For odd $q$, we find that the unique mFTI is given by the product of the $1/q$ Laughlin state and its time-reversal conjugate. The minimally-charged anyon has an electric charge $e^\ast=e/q$. The squared total quantum dimension is $\cD^2 = 2q^2$. 

It is also interesting to consider the stability of the gapless states at the interface between different mFTIs, based on which one can deduce a full classification of all the mFTIs with $\sigma^{\rm sh}=\frac{1}{q}$. For convenience, we denote the mFTI with $\sigma^{\rm sh}=\frac{1}{q}$ as mFTI$_q$. For our classification, if mFTI$_q$ and mFTI$_{q'}$ can admit a gapped interface that respects charge conservation and time-reversal symmetry, we consider them as equivalent, i.e. ${\rm mFTI}_q\sim{\rm mFTI}_{q'}$. Note that we allow the breaking of the $S_z$ conservation on the interface as far as this equivalence relation is concerned. Asking whether mFTI$_q$ and mFTI$_{q'}$ have a stable gapless interface is equivalent to asking whether a system obtained from stacking ${\rm mFTI}_q$ and ${\rm mFTI}_{q'}$ has a stable gapless edge state. It is helpful to write $q= 2^k p$ with $k= 0, 1,2,3,...$ and $p$ odd. It turns out that each $k$ corresponds to a distinct class of mFTIs, which we show in the following. 

Consider mFTI$_{q_1}$ and mFTI$_{q_2}$ with $q_1=2^{k_1} p_1$ and $q_2 = 2^{k_2} {p_2}$, where $p_1$ and $p_2$ are odd integers. 
 Without loss of generality, let us assume $k_1\geq k_2$. When we stack mFTI$_{q_1}$ and mFTI$_{q_2}$, the total spin-Hall conductance is 
\begin{equation}
  \sigma^{\rm sh}_{\rm tot} = \frac{1}{q_1} + \frac{1}{q_2}=\frac{2^{k_1-k_2} p_1 + p_2}{2^{k_1} p_1 p_2}.  
\end{equation} 
 mFTI$_{q_1}$ (mFTI$_{q_2}$) by itself has anyon with the minimal electric charge $\frac{e}{2^{k_1} p_1}$ ($\frac{e}{2^{k_2} p_2}$). Hence, the stacked system admits a minimal charge of 
\begin{equation}
  e^\ast= \frac{{\rm gcd}(2^{k_1-k_2} p_1 , p_2)}{2^{k_1} p_1 p_2} e = \frac{{\rm gcd}(p_1,p_2)}{2^{k_1} p_1 p_2} e.
\end{equation} 
Knowing that $p_{1,2}$ are odd, we can conclude that $\sigma^{\rm sh}_{\rm tot} /e^\ast$ is odd when $k_1 \neq k_2$ and even when $k_1 = k_2$. By the criteria obtained in Ref. \onlinecite{FTI}, when $\sigma^{\rm sh}_{\rm tot} /e^\ast$ is odd, i.e. $k_1 \neq k_2$, the stacked system has a stable gapless edge protected by charge conservation and time-reversal symmetry. When $\sigma^{\rm sh}_{\rm tot} /e^\ast$ is even, i.e. $k_1 = k_2$, the stacked system admits a charge-conserving, time-reversal-symmetric gapped edge~\cite{FTI2}. Therefore, the mFTI$_q$ with $q=2^k p$ is classified by the non-negative integer $k$ according to the equivalence relation defined above. For each class, we can choose the state mFTI$_{q=2^{k}}$ as the representative.

\section{Summary and Discussion}

Motivated by a recent discovery of the fractional quantum spin Hall effect in twisted bilayer MoTe$_2$\cite{FTIex}, in this work, we proposed a minimal fractional topological insulator that can potentially be realized in a pair of conjugate Chern bands, both at electron filling $\nu = 1/2$. The mFTI we proposed has the minimal total quantum dimension amongst all the topological orders that can be potentially realized in the same setting (especially with the same symmetry and the fractionally quantized spin-Hall conductance $\sigma^{\rm sh} = 1/2$ suggested by the experiment on twisted bilayer MoTe$_2$). 

We found that the minimally-charged anyon in mFTI has a fractional charge $e^\ast = e/2$, which is different from all product TOs. The gapless edge state of this mFTI is protected by charge conservation and time-reversal symmetry even when $S_z$ is not conserved. We also showed that the mFTI is the common descendant of various product TOs of the two valleys, which have larger quantum dimensions. The mFTI is also the common symmetry-enriched TO that can be promoted from different bosonic SPT states by gauging the discrete subgroup of their symmetries.

We caution that a more thorough analysis, similar to that for the $\nu = 2/3$ fractional quantum state\cite{kanefisherpol,kanefisher1,kanefisher2}, is required to understand the quantization of the transport signals measured in the 4-terminal geometry considered in experiment Ref. \onlinecite{FTIex}. In particular, to compare our mFTI state with this experiment, we need to carefully examine the interactions between the edge modes, disorder, and the coupling between the edge modes and the electrical contact, which would be an interesting topic for future studies.

When the constraint on time-reversal symmetry is released (while maintaining the Hall transports requirements associated with the electric charge and $S_z$), we found mFTI belonging to an 8-fold classified list of minimal TOs. The mFTI and 3 other minimal TOs are Abelian topological orders, while the remaining 4 are non-Abelian. All 7 minimal TOs other than the mFTI exhibit quantized thermal Hall effects in addition to the required Hall responses
associated with the charge and $S_z$ quantum numbers. In particular, it is worth noting that the simplest non-Abelian minimal TOs must break the time-reversal symmetries.

We generalized the construction of mFTI (and the minimal TOs) to a pair of conjugate $1/q$-filled Chern bands (for positive integer $q$). We classify all the mFTIs via the stability of the gapless interfaces between them.

It is interesting to explore the mFTI's and the minimal TOs' nearby phases and transitions in future research. For example, condensing the ${\bf e}$ anyon in the mFTI can result in superconductivity, while the condensation of ${\bf m}$ can result in a state with a spontaneously broken $S_z$ symmetry, equivalent to an inter-valley coherent state (due to spin-valley locking in the moir\'{e} TMD systems). It would be interesting to investigate the nature of these transitions and the details of the resulting phases. We note that recently both a superconductor~\cite{TMDSC,TMDSC2} and an insulator~\cite{TMDSC} were observed in a pair of conjugate Chern bands at half-filling, similar to our set-up, and the insulator phase may indeed have inter-valley coherent spin order~\cite{Devakul_2021}. It would be interesting to explore the detailed relation between the mFTI discussed in this work and the observed superconductivity and insulating phase. Another direction is to explore possible compressible phases in the vicinity of the mFTI and minimal TOs. Examples of such compressible phases include the conjugate composite Fermi liquid \cite{nayanccfl} and the vortex spin liquid \cite{zhangccfl}.  Investigating the universal properties of the transitions between the compressible states and the minimal TOs would be an interesting topic for future studies.

In this work, we've identified the minimal TOs under the symmetry and spin-Hall transport requirements. They have the smallest possible total quantum dimensions. And they are different from the product TOs. It would also be interesting to investigate if there are other possible topological state candidates satisfying the same symmetry and transport requirements that are simpler, namely smaller in the total quantum dimensions, than the product TOs.

%At half-filling of a pair of conjugate Landau levels (or Chern bands), the picture of flux attachment could naturally lead to a different state, i.e. the conjugate composite Fermi liquid (cCFL) state discussed recently~\cite{zhangccfl1,nayanccfl,zhangccfl2,shi2024excitonic}.In fact, despite the compressible bulk, the cCFL may potentially also generate a similar signal as what was observed experimentally. The electron Green's function of the CFL is short-ranged in the bulk. But it decays as a power-law at the edge~\cite{cfledge1,cfledge2,cfledge3,cfledge4}. These ``edge states", though not as sharply defined as in a fully gapped TO, still lead to the quantized Hall conductivity at $\nu = 1/2$ (without plateau) of the fractional quantum Hall systems. The bulk of the cCFL is a ``bad metal" whose longitudinal conductivity is $\sigma_{xx} \sim 1/\sigma^\mathrm{cf}$, where $\sigma^{\mathrm{cf}}$ is the conductivity of the composite fermions. Hence, when the composite fermions form a good conductor with a large $\sigma^{\mathrm{cf}}$, the charge transport could be dominated (or short-circuited) by the edge states, depending on the size and geometry of the sample. A measurement of the bulk charge compressibility can clearly distinguish the cCFL and a fully gapped FTI.

{\it Acknowledgement} - The authors thank Leon Balents and Matthew Fisher for very helpful discussions. C.X. is supported by the Simons foundation through the Simons investigator program. M. C. acknowledges support from NSF under Award No. DMR-1846109, and is currently supported in part by NSF under Award No. DMR-2424315. C.-M.J. is supported by the Alfred P. Sloan Foundation through a Sloan
Research Fellowship. 

While finishing this paper, we became aware of another independent work~\cite{Julian} studying the edge state of FTIs that are a product of conjugate TOs from two valleys, which, as we discussed, should have a larger quantum dimension than the mFTI constructed in this paper. Shortly after our work was posted on Arxiv, an independent work\cite{ZhangFQSH} studying the possible $p$-wave pairings on the neutral Fermi surface of the vortex spin liquid appeared. The resulting phases turn out to be topologically equivalent to our $\mathbb{Z}_4$ mFTI and the non-Abelian minimal TO discussed in Sec. \ref{typeII}.

\appendix

\section{Classifying minimal TOs with or without time-reversal symmetry}

In this appendix, we perform a systematic classification of minimal TOs for gapped topological states of electrons with fractional spin-Hall conductance $\sigma^{\rm sh}=\frac{1}{q}$ (with positive integer $q$), but no diagonal charge or spin quantum Hall effect (i.e. no transverse spin current responding to spin gauge field). This classification of minimal TOs only assumes the charge and the spin-$S_z$ conservations. 
Then, we identify the mFTIs by further imposing the time-reversal symmetry ${\cal T}$. It turns out that there is a unique mFTI for each $q$. We separately discuss the cases with even and odd $q$'s. For even $q$, we find that there are exactly eight minimal TOs, all with squared total quantum dimension $\mathcal{D}^2=8q^2$. One of them is the $\Z_{2q}$ TO, the same TO as the $\Z_{2q}$ toric code.
Further imposing time-reversal symmetry, we find that this $\Z_{2q}$ TO is singled out as the unique mFTI that satisfies all the symmetry and transport requirements. For odd $q$, there is a unique minimal TO given by a product TO $\U(1)_q\boxtimes \U(1)_{-q}$, which can also be compatible with time-reversal symmetry. 

\subsection{Minimal TOs without assuming time-reversal symmetry}
Let's first discuss the case with even $q$. Following the notations in the main text, define $v_{1,2}$ as the anyons generated by the 2$\pi$ fluxes 
%vision 
of $\U(1)_{\uparrow, \downarrow}$. We denote the $\U(1)_{\uparrow, \downarrow}$-charges of an anyon $x$ by $q_x = (q_{x,1}, q_{x,2})$. The charge and spin Hall responses translate into the following conditions on the self and mutual statistics of $v_{1,2}$: 
\begin{equation}
    \theta_{v_1} = e^{\frac{\ii\pi}{q}}, \theta_{v_2}=e^{-\frac{\ii\pi}{q}}, M_{v_1, v_2}=1.
\end{equation}
Note that $v_1$ has charge $q_{v_1}=(\frac{1}{q},0)$ and $v_2$ has $q_{v_2}=(0,-\frac{1}{q})$.

We see that $v_1^q v_2^q$ is a boson carrying charge $(1,-1)$. Similarly, the boson $v_1^{2q}$ ($v_2^{2q}$) has charge $(2,0)$ ($(0,2)$).  Their charges can be neutralized by attaching physical bosons (Cooper pairs or excitons). Thus, we can always condense $v_1^q v_2^q, v_1^{2q}$ and $v_2^{2q}$ to reduce the TO, without breaking the $\U(1)_\uparrow\times \U(1)_\downarrow$ symmetry. 
After the condensation, we find that the TO contains the subcategory $\cC_0$ generated by $v_1, v_2$ subject to the relations $v_1^q=v_2^q=b,~b^2=1$. Here, we've introduced the notation $b$, which labels a self-boson with a $\Z_2$ fusion rule. In addition, we have 
\begin{equation}
    M_{b, v_1}=M_{v_1^q, v_1}=\theta_{v_1}^{2q}=1, \nonumber
    M_{b, v_2}=M_{v_2^q, v_2}=\theta_{v_2}^{2q}=1.
\end{equation}
If we further condense $v_{1}^q$ and $v_{2}^q$, the resulting TO is the product of the $\U(1)_q$ TO generated by $v_1$ and the $\U(1)_{-q}$ TO generated by $v_2$. Therefore, we conclude that $\mathcal{C}_0$ describes the ``symmetrization" (or ``equivariantization") of the $\U(1)_q\boxtimes\U(1)_{-q}$ TO enriched by a $\Z_2$ symmetry. The $b=v_1^q=v_2^q$ anyon is identified as the $\Z_2$ charge. We infer from the fusion rule that both $v_1$ and $v_2$ carry ``$1/q$ charge" under this $\Z_2$ symmetry.

To find the minimal TO, we use the fact that the TO is a (fermionic) modular extension of $\cC_0$. Using a theorem in Ref. \onlinecite{Muger_2003} (Proposition 5.1), the total quantum dimension of any modular extension satisfies $\mathcal{D}^2\geq 8q^2$ (counting the electron as a nontrivial anyon type). In this case, because $\cC_0$ has a $\Z_2$ transparent center (generated by $b$), all minimal modular extensions are obtained from gauging the $\Z_2$ symmetry. One of the modular extensions is just the $\Z_{2q}$ toric code $\mathrm{D}(\Z_{2q})$.
However, when gauging the $\Z_2$ symmetry, one also has additional freedom in choosing a topological term for the $\Z_2$ gauge field, or in other words, stacking a $\Z_2$ SPT state. Notice that we are considering a fermionic system. So, more precisely, the stacked SPT state should be a $\Z_2\times \Z_2^f$ fermionic SPT state, which is classified by $\Z_8$. %Formally, they can all 
Hence, all possible fermionic modular extensions of $\cC_0$ can
be represented as follows:
\begin{equation}
    \mathcal{M}_n=\frac{\mathrm{D}(\Z_{2q})\boxtimes \mathrm{Spin}(n)_1}{({\bf e}^q{\bf m}^q, \psi, f)}, n=0,1,\dots, 7.
    \label{eq:MTO_evenq}
\end{equation}
Here, Spin$(n)_1$ denotes the TO of the level-1 Spin$(n)$ Chern-Simons theory, $\psi$ is the neutral fermion in Spin$(n)_1$, and $f$ represents the physical electron/hole. ${\bf e}$ and ${\bf m}$ are the two elementary anyons in the $\Z_{2q}$ toric code $\mathrm{D}(\Z_{2q})$. The quotient means condensing the bound state of ${\bf e}^q{\bf m}^q$, $\psi$, and $f$. The reason why we only have $n=0,1,2,..,7$ is that ${\cal M}_n$ and ${\cal M}_{n+8}$ share the same anyon content up to a relabeling of anyon types. 

One can check that all the eight theories above are distinct from each other. Without further conditions, we have found all possible minimal TOs. All the minimal TOs share the same minimal electric charge $e^* = e/q$. Notice that even (odd) $n$'s correspond to the type-I (type-II) minimal TOs. For example, the non-Abelian TO discussed in Sec. \ref{typeII} corresponds to $\mathcal{M}_1$ for $q=2$. 

In the case of $q=2$, the type-I minimal TOs ${\cal M}_{0,2,4,6}$ and the type-II minimal TOs ${\cal M}_{1,3,5,7}$ follow the braiding statistics listed in Sec. \ref{bootstrap}. The extra data of the topological spin of the anyon $y$ can be obtained from Eq. \eqref{eq:MTO_evenq}: 
\begin{align}
    \theta_y = e^{\ii n\pi/8}.
    \label{eq:thetay}
\end{align}
Also, recall that $y$ is an Abelian anyon in ${\cal M}_{0,2,4,6}$ and a non-Abelian anyon in ${\cal M}_{1,3,5,7}$. From the value of $\theta_y$ and the fractional charge $q_y= (1/4,1/4)$, we can obtain the following fusion rule:
\begin{align}
    y \times y = \begin{cases}
    v_1 v_2^3, & n =0,4, \\
    v_1 v_2 f,& n= 2,6, \\
    v_1 v_2^3 + v_1 v_2 f,& n= 1,3,5,7.
    \end{cases}
    \label{eq:y_fusion}
\end{align}

The chiral central charge of these minimal TOs requires more careful treatment. Ref. \onlinecite{Lapa2019Indicator} showed that, in a charge-conserved TO built only from charge-$e$ fermions, the chiral central charge $c$ can be determined by the anyon data of the TO and its charge fractionalization modulo 8. While ${\cal M}_n$ and ${\cal M}_{n+8}$ share the same anyon data, their charge fractionalization can still be different. A manifestation of such difference is that the topological spin $\theta_y$ in Eq. \eqref{eq:thetay} of the anyon $y$ with fractional charge $(1/4,1/4)$ acquires an extra $-1$ factor when we change $n$ to $n+8$. If one further attaches an $f$ to $y$ to restore the topological spin of $y$, the fractional charge of $y$ is then altered. Hence, at the level of the topological order alone, the minimal TOs are classified by ${\cal M}_n$ with $n=0,1,...,7$. As minimal TOs with $\U(1)_\uparrow\times \U(1)_\downarrow$ symmetry fractionalization, we should allow the range of $n$ to be $n=0,1,2,...,15$. These 16 TOs can distinguished by $\theta_y$ (though the fusion rule of $y$ still only depends on $n$ mod 8 via Eq. \eqref{eq:y_fusion}), with $q_y=(1/4,1/4)$. The chiral central charge of these TOs ${\cal M}_{n=0,1,...,15}$ are given by
\begin{align}
    c = n/2~~{\rm mod}~8.
    \label{eq:ccc}
\end{align}
Here the mod 8 chiral central charge is because one can stack a $E_8$ state formed out of neutral bosons, without changing the bulk SET order.

For $q$ odd, we note that $v_1^q$ is actually a fermion carrying charges $(1,0)$. We can thus identify it with the spin-up electron. Similarly, we can identify $v_2^q$ with the spin-down hole. With this identification, we obtain the minimal TO
$\U(1)_q\boxtimes \U(1)_{-q}$, which has same TO as the $\Z_q$ toric code (in a fermionic system). This state is nothing but the simple fractional QSH state discussed in Ref. \onlinecite{FTI} whose $K$-matrix and charge vectors are given by
\begin{equation}
    K=\begin{pmatrix}
        q & 0\\
        0 & -q
    \end{pmatrix},~t_\uparrow=\begin{pmatrix}
        1  \\
        0
    \end{pmatrix},~t_\downarrow = \begin{pmatrix}
        0 \\
        1
    \end{pmatrix}.
\end{equation}
This TO has the minimal squared total quantum dimension $\cD^2 = 2q^2$. When $q=1$, this $K$-matrix theory above reduces to the theory of the non-interacting topological insulator.

\subsection{Imposing time-reversal symmetry}
For odd $q$, the unique minimal TO is compatible with time-reversal symmetry. Hence, the unique minimal TO $\U(1)_q\boxtimes \U(1)_{-q}$ is the unique mFTI.

For the rest of the discussion, we focus on even $q$. From the chiral central charge Eq. \eqref{eq:ccc}, only $n=0$ is compatible with time-reversal symmetry. The theory ${\cal M}_0$ is the mFTI discussed in Sec. \ref{sec:mFTI_construction} and Sec. \ref{sec:mFTI_q}. Hence, the mFTIs with $\mathbb{Z}_{2q}$ TOs are the {\it unique} mFTIs. 

For even $n$, we can provide a different analysis that also confirms that ${\cal M}_0$ is the only minimal TO compatible with time-reversal symmetry. 
With $n$ even, $\mathcal{M}_n$ is an Abelian TO and can be described by a $K$-matrix theory. First, recall that the $\U(1)_q\boxtimes \U(1)_{-q}$ theory is described by the following Chern-Simons theory: 
\begin{equation}
\mathcal{L}=\frac{\ii q}{4\pi}a_1da_1-\frac{\ii q}{4\pi}a_2da_2 + \frac{\ii }{2\pi}Ad(a_1+a_2).
\label{eqn:U1q}
\end{equation}
$v_1$ ($v_2$) carries a unit gauge charge under $a_1$($a_2$). 
To describe the $\Z_2$ symmetry enrichment, it is convenient to first enlarge the $\Z_2$ symmetry to $\U(1)$, under which $v_1$ and $v_2$ both carry charge $1/q$. $A$ in Eq. \eqref{eqn:U1q} denotes the background gauge field for the enlarged U(1) symmetry. Then, we Higgs the U(1) to $\Z_2$, which can be implemented by the BF term $\frac{1}{\pi}BdA$ with a U(1) gauge field $B$. We can then promote both $A$ and $B$ to dynamical gauge fields to gauge the symmetry. The SPT layer can be accounted for by adding a Chern-Simons term $-\frac{n}{8\pi}AdA$ for $A$. All together, we have found the following $K$-matrix (gauge fields ordered as $a_1,a_2, A, B$):
\begin{equation}
    K_{\mathcal{M}_n}=\begin{pmatrix}
        q & 0 & 1 & 0\\
        0 & -q & 1 & 0\\
        1 & 1 & -n/2 & 2\\
        0 & 0 & 2 & 0
    \end{pmatrix}.
\end{equation}
Evidently, we have the following identification: $v_1=(1,0,0,0), v_2=(0,1,0,0)$. In addition, ${\bf m}=(0,0,0,-1)$ represents a $\Z_2$ gauge flux anyon.
The fusion rule of the Abelian anyons in this theory is given by the anyon group
$\Z_{2q}\times\Z_{2q}$, the same as the $\Z_{2q}$ toric code. 
%For $q=2$, this TO is listed as the $F_4$ theory in  Ref. \onlinecite{Wang:2020nmz}.
The two generators of the anyon group can be chosen as $v_1=(1,0,0,0)$ and ${\bf m}=(0,0,0,-1)$. And we have $v_2=v_1^{-1+nq/2}{\bf m}^{2}$. The self and mutual statistics of these generators are given by
\begin{equation}
    \theta_{v_1}=e^{\frac{\ii\pi}{q}},~\theta_{\bf m}=e^{\frac{\ii\pi n}{8}},~ M_{v_1,{\bf m}}=e^{\frac{\ii\pi}{q}}.
\end{equation}

We now determine the time-reversal transformations of anyons. First, since $v_1$ and $v_2$ are generated by $2\pi$ fluxes of $\U(1)_{\uparrow,\downarrow}$, they should have the following transformations:
\begin{equation}
     \cT(v_1)=v_2^{-1}=v_1^{1-nq/2} {\bf m}^{-2}, \cT(v_2)=v_1^{-1}.
\end{equation}
We can then use the exchange and braiding statistics to fix the transformation of ${\bf m}$. In general, we may write $\cT({\bf m})={\bf m}^a v_1^b f^c$ with $a,b,c \in \Z$. $M_{\cT({\bf m}), \cT(v_1)}=e^{-\frac{\ii\pi}{q}}$ yields $a=-1$. Then, the requirement $\theta_{\cT({\bf m})} = \theta_{{\bf m}}^*$ yields
\begin{equation}
    \theta_{\cT({\bf m})} = e^{\ii\pi \left(\frac{n}{8}+\frac{b(b-1)}{q} \right)}(-1)^c = e^{-\frac{\ii\pi n}{8}} = \theta_{{\bf m}}^*.
    \label{eq:theta_m_TR_constraint}
\end{equation}

%It turns out that there are two options:
%\begin{equation}
%    \cT({\bf m})=v_1^q {\bf m}^{-1}\text{ or }\cT({\bf m})={\bf m}^{-1}f,
%\end{equation}
%where $f$ is the physical electron. Therefore, the theory is time-reversal-invariant with $\cT^2=1$. 

There are extra constraints coming from the compatibility between time-reversal symmetry and the $\U(1)_\uparrow\times \U(1)_\downarrow$ symmetry. More specifically, the $\U(1)_{\uparrow, \downarrow}$ charges should be inter-changed under $\cT$. That means the time-reversal partner ${\cal T}(x)$ of the anyon $x$ should carry charges:
\begin{align}
    q_{\cT(x)} = (q_{x,2}~,~q_{x,1} ) + (n_1, n_2)
    %~{\rm with~} n_1+ n_2\equiv 0\text{ mod }2.
    \label{eq:TR_charge_constraint}
\end{align}
% \begin{equation}
% \begin{split}
%     q_{\cT(x), 1}&= q_{x,2}+n_1,\\
%     q_{\cT(x), 2}&= q_{x,1}+n_2,\\
%     n_1+&n_2\equiv 0\text{ mod }2.
% \end{split}
% \end{equation}
% Here $n_1$ and $n_2$ represent local bosons.
Here, $(n_1, n_2)$ represents the charge of local bosons, i.e. $n_{1,2}\in\Z$, and $n_1+ n_2\equiv 0\text{ mod }2$.

Combining $\cT({\bf m})={\bf m}^{-1} v_1^b f^c$ with the constrain Eq. \eqref{eq:TR_charge_constraint}, we have
\begin{equation}
\begin{split}
     &-q_{{\bf m},1}+\frac{b}{q}+c=q_{{\bf m},2}+n_1,\\ 
     &-q_{{\bf m},2}=q_{{\bf m},1}+n_2,
\end{split}
\end{equation}
which leads to
\begin{equation}
    -q_{{\bf m},1}-q_{{\bf m},2}=n_1-\frac{b}{q}-c=n_2.
\end{equation}
Therefore, $b$ must be an integer multiple of $q$. Since $b$ is defined modulo $2q$, we only have the options $b=0$ or $b=q$.

For $b=0$, we find $n=0$ or $4$ are the only possibilities compatible with Eq. \eqref{eq:theta_m_TR_constraint}. For $n=4$, it follows that $c=1$ and $\cT({\bf m})={\bf m}^{-1}f$. 

For $b=q$, we again find $n=0$ or $4$ to be the only possibilities compatible with Eq. \eqref{eq:theta_m_TR_constraint}. For $n=4$, it follows that $c=0$, and we have $\cT({\bf m})={\bf m}^{-1}v_1^q$.

At this point, we only need to examine ${\cal M}_4$. With $n=4$, for both $b=0$ and $b=q$, we find $n_1-n_2=\frac{b}{q}+c=1$, contradicting the requirement that $n_1+n_2$ is even. We conclude that it is impossible for the theory ${\cal M}_4$ to have a charge assignment under $\U(1)_\uparrow\times \U(1)_\downarrow$ that is compatible with the time-reversal symmetry. Hence, ${\cal M}_0$ is the unique minimal TO that is consistent with time-reversal symmetry. Hence, it is the unique mFTI.

%For the $\mathcal{M}_4$ theory, we can fully determine the time-reversal transformations of the anyons: 
%\begin{equation}
%    \cT(v_1)=v_2^{-1}=v_1 {\bf m}^{-2},\: \cT({\bf m})=v_1^q {\bf m}^{-1}. 
%\end{equation}

\bibliography{TO}

%merlin.mbs apsrev4-1.bst 2010-07-25 4.21a (PWD, AO, DPC) hacked
%Control: key (0)
%Control: author (0) dotless jnrlst
%Control: editor formatted (1) identically to author
%Control: production of article title (0) allowed
%Control: page (1) range
%Control: year (0) verbatim
%Control: production of eprint (0) enabled
\begin{thebibliography}{88}%
\makeatletter
\providecommand \@ifxundefined [1]{%
 \@ifx{#1\undefined}
}%
\providecommand \@ifnum [1]{%
 \ifnum #1\expandafter \@firstoftwo
 \else \expandafter \@secondoftwo
 \fi
}%
\providecommand \@ifx [1]{%
 \ifx #1\expandafter \@firstoftwo
 \else \expandafter \@secondoftwo
 \fi
}%
\providecommand \natexlab [1]{#1}%
\providecommand \enquote  [1]{``#1''}%
\providecommand \bibnamefont  [1]{#1}%
\providecommand \bibfnamefont [1]{#1}%
\providecommand \citenamefont [1]{#1}%
\providecommand \href@noop [0]{\@secondoftwo}%
\providecommand \href [0]{\begingroup \@sanitize@url \@href}%
\providecommand \@href[1]{\@@startlink{#1}\@@href}%
\providecommand \@@href[1]{\endgroup#1\@@endlink}%
\providecommand \@sanitize@url [0]{\catcode `\\12\catcode `\$12\catcode `\&12\catcode `\#12\catcode `\^12\catcode `\_12\catcode `\%12\relax}%
\providecommand \@@startlink[1]{}%
\providecommand \@@endlink[0]{}%
\providecommand \url  [0]{\begingroup\@sanitize@url \@url }%
\providecommand \@url [1]{\endgroup\@href {#1}{\urlprefix }}%
\providecommand \urlprefix  [0]{URL }%
\providecommand \Eprint [0]{\href }%
\providecommand \doibase [0]{http://dx.doi.org/}%
\providecommand \selectlanguage [0]{\@gobble}%
\providecommand \bibinfo  [0]{\@secondoftwo}%
\providecommand \bibfield  [0]{\@secondoftwo}%
\providecommand \translation [1]{[#1]}%
\providecommand \BibitemOpen [0]{}%
\providecommand \bibitemStop [0]{}%
\providecommand \bibitemNoStop [0]{.\EOS\space}%
\providecommand \EOS [0]{\spacefactor3000\relax}%
\providecommand \BibitemShut  [1]{\csname bibitem#1\endcsname}%
\let\auto@bib@innerbib\@empty
%</preamble>
\bibitem [{\citenamefont {Spanton}\ \emph {et~al.}(2018)\citenamefont {Spanton}, \citenamefont {Zibrov}, \citenamefont {Zhou}, \citenamefont {Taniguchi}, \citenamefont {Watanabe}, \citenamefont {Zaletel},\ and\ \citenamefont {Young}}]{youngFCI}%
  \BibitemOpen
  \bibfield  {author} {\bibinfo {author} {\bibfnamefont {Eric~M.}\ \bibnamefont {Spanton}}, \bibinfo {author} {\bibfnamefont {Alexander~A.}\ \bibnamefont {Zibrov}}, \bibinfo {author} {\bibfnamefont {Haoxin}\ \bibnamefont {Zhou}}, \bibinfo {author} {\bibfnamefont {Takashi}\ \bibnamefont {Taniguchi}}, \bibinfo {author} {\bibfnamefont {Kenji}\ \bibnamefont {Watanabe}}, \bibinfo {author} {\bibfnamefont {Michael~P.}\ \bibnamefont {Zaletel}}, \ and\ \bibinfo {author} {\bibfnamefont {Andrea~F.}\ \bibnamefont {Young}},\ }\bibfield  {title} {\enquote {\bibinfo {title} {Observation of fractional chern insulators in a van der waals heterostructure},}\ }\href {\doibase 10.1126/science.aan8458} {\bibfield  {journal} {\bibinfo  {journal} {Science}\ }\textbf {\bibinfo {volume} {360}},\ \bibinfo {pages} {62--66} (\bibinfo {year} {2018})}\BibitemShut {NoStop}%
\bibitem [{\citenamefont {{Xie}}\ \emph {et~al.}(2021)\citenamefont {{Xie}}, \citenamefont {{Pierce}}, \citenamefont {{Park}}, \citenamefont {{Parker}}, \citenamefont {{Khalaf}}, \citenamefont {{Ledwith}}, \citenamefont {{Cao}}, \citenamefont {{Lee}}, \citenamefont {{Chen}}, \citenamefont {{Forrester}}, \citenamefont {{Watanabe}}, \citenamefont {{Taniguchi}}, \citenamefont {{Vishwanath}}, \citenamefont {{Jarillo-Herrero}},\ and\ \citenamefont {{Yacoby}}}]{harvardFCI}%
  \BibitemOpen
  \bibfield  {author} {\bibinfo {author} {\bibfnamefont {Yonglong}\ \bibnamefont {{Xie}}}, \bibinfo {author} {\bibfnamefont {Andrew~T.}\ \bibnamefont {{Pierce}}}, \bibinfo {author} {\bibfnamefont {Jeong~Min}\ \bibnamefont {{Park}}}, \bibinfo {author} {\bibfnamefont {Daniel~E.}\ \bibnamefont {{Parker}}}, \bibinfo {author} {\bibfnamefont {Eslam}\ \bibnamefont {{Khalaf}}}, \bibinfo {author} {\bibfnamefont {Patrick}\ \bibnamefont {{Ledwith}}}, \bibinfo {author} {\bibfnamefont {Yuan}\ \bibnamefont {{Cao}}}, \bibinfo {author} {\bibfnamefont {Seung~Hwan}\ \bibnamefont {{Lee}}}, \bibinfo {author} {\bibfnamefont {Shaowen}\ \bibnamefont {{Chen}}}, \bibinfo {author} {\bibfnamefont {Patrick~R.}\ \bibnamefont {{Forrester}}}, \bibinfo {author} {\bibfnamefont {Kenji}\ \bibnamefont {{Watanabe}}}, \bibinfo {author} {\bibfnamefont {Takashi}\ \bibnamefont {{Taniguchi}}}, \bibinfo {author} {\bibfnamefont {Ashvin}\ \bibnamefont {{Vishwanath}}}, \bibinfo {author} {\bibfnamefont {Pablo}\ \bibnamefont {{Jarillo-Herrero}}}, \ and\
  \bibinfo {author} {\bibfnamefont {Amir}\ \bibnamefont {{Yacoby}}},\ }\bibfield  {title} {\enquote {\bibinfo {title} {{Fractional Chern insulators in magic-angle twisted bilayer graphene}},}\ }\href {\doibase 10.1038/s41586-021-04002-3} {\bibfield  {journal} {\bibinfo  {journal} {\nat}\ }\textbf {\bibinfo {volume} {600}},\ \bibinfo {pages} {439--443} (\bibinfo {year} {2021})},\ \Eprint {http://arxiv.org/abs/2107.10854} {arXiv:2107.10854 [cond-mat.mes-hall]} \BibitemShut {NoStop}%
\bibitem [{\citenamefont {Cai}\ \emph {et~al.}(2023)\citenamefont {Cai}, \citenamefont {Anderson}, \citenamefont {Wang}, \citenamefont {Zhang}, \citenamefont {Liu}, \citenamefont {Holtzmann}, \citenamefont {Zhang}, \citenamefont {Fan}, \citenamefont {Taniguchi}, \citenamefont {Watanabe}, \citenamefont {Ran}, \citenamefont {Cao}, \citenamefont {Fu}, \citenamefont {Xiao}, \citenamefont {Yao},\ and\ \citenamefont {Xu}}]{xuFQAH1}%
  \BibitemOpen
  \bibfield  {author} {\bibinfo {author} {\bibfnamefont {Jiaqi}\ \bibnamefont {Cai}}, \bibinfo {author} {\bibfnamefont {Eric}\ \bibnamefont {Anderson}}, \bibinfo {author} {\bibfnamefont {Chong}\ \bibnamefont {Wang}}, \bibinfo {author} {\bibfnamefont {Xiaowei}\ \bibnamefont {Zhang}}, \bibinfo {author} {\bibfnamefont {Xiaoyu}\ \bibnamefont {Liu}}, \bibinfo {author} {\bibfnamefont {William}\ \bibnamefont {Holtzmann}}, \bibinfo {author} {\bibfnamefont {Yinong}\ \bibnamefont {Zhang}}, \bibinfo {author} {\bibfnamefont {Fengren}\ \bibnamefont {Fan}}, \bibinfo {author} {\bibfnamefont {Takashi}\ \bibnamefont {Taniguchi}}, \bibinfo {author} {\bibfnamefont {Kenji}\ \bibnamefont {Watanabe}}, \bibinfo {author} {\bibfnamefont {Ying}\ \bibnamefont {Ran}}, \bibinfo {author} {\bibfnamefont {Ting}\ \bibnamefont {Cao}}, \bibinfo {author} {\bibfnamefont {Liang}\ \bibnamefont {Fu}}, \bibinfo {author} {\bibfnamefont {Di}~\bibnamefont {Xiao}}, \bibinfo {author} {\bibfnamefont {Wang}\ \bibnamefont {Yao}}, \ and\ \bibinfo {author}
  {\bibfnamefont {Xiaodong}\ \bibnamefont {Xu}},\ }\bibfield  {title} {\enquote {\bibinfo {title} {Signatures of fractional quantum anomalous hall states in twisted {MoTe}2},}\ }\href {\doibase 10.1038/s41586-023-06289-w} {\bibfield  {journal} {\bibinfo  {journal} {Nature}\ }\textbf {\bibinfo {volume} {622}},\ \bibinfo {pages} {63--68} (\bibinfo {year} {2023})}\BibitemShut {NoStop}%
\bibitem [{\citenamefont {Park}\ \emph {et~al.}(2023)\citenamefont {Park}, \citenamefont {Cai}, \citenamefont {Anderson}, \citenamefont {Zhang}, \citenamefont {Zhu}, \citenamefont {Liu}, \citenamefont {Wang}, \citenamefont {Holtzmann}, \citenamefont {Hu}, \citenamefont {Liu}, \citenamefont {Taniguchi}, \citenamefont {Watanabe}, \citenamefont {Chu}, \citenamefont {Cao}, \citenamefont {Fu}, \citenamefont {Yao}, \citenamefont {Chang}, \citenamefont {Cobden}, \citenamefont {Xiao},\ and\ \citenamefont {Xu}}]{xuFQAH2}%
  \BibitemOpen
  \bibfield  {author} {\bibinfo {author} {\bibfnamefont {Heonjoon}\ \bibnamefont {Park}}, \bibinfo {author} {\bibfnamefont {Jiaqi}\ \bibnamefont {Cai}}, \bibinfo {author} {\bibfnamefont {Eric}\ \bibnamefont {Anderson}}, \bibinfo {author} {\bibfnamefont {Yinong}\ \bibnamefont {Zhang}}, \bibinfo {author} {\bibfnamefont {Jiayi}\ \bibnamefont {Zhu}}, \bibinfo {author} {\bibfnamefont {Xiaoyu}\ \bibnamefont {Liu}}, \bibinfo {author} {\bibfnamefont {Chong}\ \bibnamefont {Wang}}, \bibinfo {author} {\bibfnamefont {William}\ \bibnamefont {Holtzmann}}, \bibinfo {author} {\bibfnamefont {Chaowei}\ \bibnamefont {Hu}}, \bibinfo {author} {\bibfnamefont {Zhaoyu}\ \bibnamefont {Liu}}, \bibinfo {author} {\bibfnamefont {Takashi}\ \bibnamefont {Taniguchi}}, \bibinfo {author} {\bibfnamefont {Kenji}\ \bibnamefont {Watanabe}}, \bibinfo {author} {\bibfnamefont {Jiun-Haw}\ \bibnamefont {Chu}}, \bibinfo {author} {\bibfnamefont {Ting}\ \bibnamefont {Cao}}, \bibinfo {author} {\bibfnamefont {Liang}\ \bibnamefont {Fu}}, \bibinfo {author}
  {\bibfnamefont {Wang}\ \bibnamefont {Yao}}, \bibinfo {author} {\bibfnamefont {Cui-Zu}\ \bibnamefont {Chang}}, \bibinfo {author} {\bibfnamefont {David}\ \bibnamefont {Cobden}}, \bibinfo {author} {\bibfnamefont {Di}~\bibnamefont {Xiao}}, \ and\ \bibinfo {author} {\bibfnamefont {Xiaodong}\ \bibnamefont {Xu}},\ }\bibfield  {title} {\enquote {\bibinfo {title} {Observation of fractionally quantized anomalous hall effect},}\ }\href {\doibase 10.1038/s41586-023-06536-0} {\bibfield  {journal} {\bibinfo  {journal} {Nature}\ }\textbf {\bibinfo {volume} {622}},\ \bibinfo {pages} {74--79} (\bibinfo {year} {2023})}\BibitemShut {NoStop}%
\bibitem [{\citenamefont {Zeng}\ \emph {et~al.}(2023)\citenamefont {Zeng}, \citenamefont {Xia}, \citenamefont {Kang}, \citenamefont {Zhu}, \citenamefont {Kn¨¹ppel}, \citenamefont {Vaswani}, \citenamefont {Watanabe}, \citenamefont {Taniguchi}, \citenamefont {Mak},\ and\ \citenamefont {Shan}}]{makFCI}%
  \BibitemOpen
  \bibfield  {author} {\bibinfo {author} {\bibfnamefont {Yihang}\ \bibnamefont {Zeng}}, \bibinfo {author} {\bibfnamefont {Zhengchao}\ \bibnamefont {Xia}}, \bibinfo {author} {\bibfnamefont {Kaifei}\ \bibnamefont {Kang}}, \bibinfo {author} {\bibfnamefont {Jiacheng}\ \bibnamefont {Zhu}}, \bibinfo {author} {\bibfnamefont {Patrick}\ \bibnamefont {Kn¨¹ppel}}, \bibinfo {author} {\bibfnamefont {Chirag}\ \bibnamefont {Vaswani}}, \bibinfo {author} {\bibfnamefont {Kenji}\ \bibnamefont {Watanabe}}, \bibinfo {author} {\bibfnamefont {Takashi}\ \bibnamefont {Taniguchi}}, \bibinfo {author} {\bibfnamefont {Kin~Fai}\ \bibnamefont {Mak}}, \ and\ \bibinfo {author} {\bibfnamefont {Jie}\ \bibnamefont {Shan}},\ }\href@noop {} {\enquote {\bibinfo {title} {Integer and fractional chern insulators in twisted bilayer mote2},}\ } (\bibinfo {year} {2023}),\ \Eprint {http://arxiv.org/abs/2305.00973} {arXiv:2305.00973 [cond-mat.mes-hall]} \BibitemShut {NoStop}%
\bibitem [{\citenamefont {Xu}\ \emph {et~al.}(2023)\citenamefont {Xu}, \citenamefont {Sun}, \citenamefont {Jia}, \citenamefont {Liu}, \citenamefont {Xu}, \citenamefont {Li}, \citenamefont {Gu}, \citenamefont {Watanabe}, \citenamefont {Taniguchi}, \citenamefont {Tong} \emph {et~al.}}]{shFQAH}%
  \BibitemOpen
  \bibfield  {author} {\bibinfo {author} {\bibfnamefont {Fan}\ \bibnamefont {Xu}}, \bibinfo {author} {\bibfnamefont {Zheng}\ \bibnamefont {Sun}}, \bibinfo {author} {\bibfnamefont {Tongtong}\ \bibnamefont {Jia}}, \bibinfo {author} {\bibfnamefont {Chang}\ \bibnamefont {Liu}}, \bibinfo {author} {\bibfnamefont {Cheng}\ \bibnamefont {Xu}}, \bibinfo {author} {\bibfnamefont {Chushan}\ \bibnamefont {Li}}, \bibinfo {author} {\bibfnamefont {Yu}~\bibnamefont {Gu}}, \bibinfo {author} {\bibfnamefont {Kenji}\ \bibnamefont {Watanabe}}, \bibinfo {author} {\bibfnamefont {Takashi}\ \bibnamefont {Taniguchi}}, \bibinfo {author} {\bibfnamefont {Bingbing}\ \bibnamefont {Tong}},  \emph {et~al.},\ }\bibfield  {title} {\enquote {\bibinfo {title} {Observation of integer and fractional quantum anomalous hall states in twisted bilayer mote2},}\ }\href@noop {} {\bibfield  {journal} {\bibinfo  {journal} {arXiv preprint arXiv:2308.06177}\ } (\bibinfo {year} {2023})}\BibitemShut {NoStop}%
\bibitem [{\citenamefont {Lu}\ \emph {et~al.}(2023)\citenamefont {Lu}, \citenamefont {Han}, \citenamefont {Yao}, \citenamefont {Reddy}, \citenamefont {Yang}, \citenamefont {Seo}, \citenamefont {Watanabe}, \citenamefont {Taniguchi}, \citenamefont {Fu},\ and\ \citenamefont {Ju}}]{juFQAH}%
  \BibitemOpen
  \bibfield  {author} {\bibinfo {author} {\bibfnamefont {Zhengguang}\ \bibnamefont {Lu}}, \bibinfo {author} {\bibfnamefont {Tonghang}\ \bibnamefont {Han}}, \bibinfo {author} {\bibfnamefont {Yuxuan}\ \bibnamefont {Yao}}, \bibinfo {author} {\bibfnamefont {Aidan~P.}\ \bibnamefont {Reddy}}, \bibinfo {author} {\bibfnamefont {Jixiang}\ \bibnamefont {Yang}}, \bibinfo {author} {\bibfnamefont {Junseok}\ \bibnamefont {Seo}}, \bibinfo {author} {\bibfnamefont {Kenji}\ \bibnamefont {Watanabe}}, \bibinfo {author} {\bibfnamefont {Takashi}\ \bibnamefont {Taniguchi}}, \bibinfo {author} {\bibfnamefont {Liang}\ \bibnamefont {Fu}}, \ and\ \bibinfo {author} {\bibfnamefont {Long}\ \bibnamefont {Ju}},\ }\href@noop {} {\enquote {\bibinfo {title} {Fractional quantum anomalous hall effect in a graphene moire superlattice},}\ } (\bibinfo {year} {2023}),\ \Eprint {http://arxiv.org/abs/2309.17436} {arXiv:2309.17436 [cond-mat.mes-hall]} \BibitemShut {NoStop}%
\bibitem [{\citenamefont {{Kang}}\ \emph {et~al.}(2024)\citenamefont {{Kang}}, \citenamefont {{Shen}}, \citenamefont {{Qiu}}, \citenamefont {{Zeng}}, \citenamefont {{Xia}}, \citenamefont {{Watanabe}}, \citenamefont {{Taniguchi}}, \citenamefont {{Shan}},\ and\ \citenamefont {{Mak}}}]{FTIex}%
  \BibitemOpen
  \bibfield  {author} {\bibinfo {author} {\bibfnamefont {Kaifei}\ \bibnamefont {{Kang}}}, \bibinfo {author} {\bibfnamefont {Bowen}\ \bibnamefont {{Shen}}}, \bibinfo {author} {\bibfnamefont {Yichen}\ \bibnamefont {{Qiu}}}, \bibinfo {author} {\bibfnamefont {Yihang}\ \bibnamefont {{Zeng}}}, \bibinfo {author} {\bibfnamefont {Zhengchao}\ \bibnamefont {{Xia}}}, \bibinfo {author} {\bibfnamefont {Kenji}\ \bibnamefont {{Watanabe}}}, \bibinfo {author} {\bibfnamefont {Takashi}\ \bibnamefont {{Taniguchi}}}, \bibinfo {author} {\bibfnamefont {Jie}\ \bibnamefont {{Shan}}}, \ and\ \bibinfo {author} {\bibfnamefont {Kin~Fai}\ \bibnamefont {{Mak}}},\ }\bibfield  {title} {\enquote {\bibinfo {title} {{Evidence of the fractional quantum spin Hall effect in moir{\'e} MoTe$_{2}$}},}\ }\href {\doibase 10.1038/s41586-024-07214-5} {\bibfield  {journal} {\bibinfo  {journal} {\nat}\ }\textbf {\bibinfo {volume} {628}},\ \bibinfo {pages} {522--526} (\bibinfo {year} {2024})},\ \Eprint {http://arxiv.org/abs/2402.03294} {arXiv:2402.03294
  [cond-mat.mes-hall]} \BibitemShut {NoStop}%
\bibitem [{\citenamefont {Wu}\ \emph {et~al.}(2019)\citenamefont {Wu}, \citenamefont {Lovorn}, \citenamefont {Tutuc}, \citenamefont {Martin},\ and\ \citenamefont {MacDonald}}]{wu2019topological}%
  \BibitemOpen
  \bibfield  {author} {\bibinfo {author} {\bibfnamefont {Fengcheng}\ \bibnamefont {Wu}}, \bibinfo {author} {\bibfnamefont {Timothy}\ \bibnamefont {Lovorn}}, \bibinfo {author} {\bibfnamefont {Emanuel}\ \bibnamefont {Tutuc}}, \bibinfo {author} {\bibfnamefont {Ivar}\ \bibnamefont {Martin}}, \ and\ \bibinfo {author} {\bibfnamefont {A.~H.}\ \bibnamefont {MacDonald}},\ }\bibfield  {title} {\enquote {\bibinfo {title} {Topological insulators in twisted transition metal dichalcogenide homobilayers},}\ }\href {\doibase 10.1103/PhysRevLett.122.086402} {\bibfield  {journal} {\bibinfo  {journal} {Phys. Rev. Lett.}\ }\textbf {\bibinfo {volume} {122}},\ \bibinfo {pages} {086402} (\bibinfo {year} {2019})}\BibitemShut {NoStop}%
\bibitem [{\citenamefont {Devakul}\ \emph {et~al.}(2021{\natexlab{a}})\citenamefont {Devakul}, \citenamefont {Cr{\'e}pel}, \citenamefont {Zhang},\ and\ \citenamefont {Fu}}]{devakul2021magic}%
  \BibitemOpen
  \bibfield  {author} {\bibinfo {author} {\bibfnamefont {Trithep}\ \bibnamefont {Devakul}}, \bibinfo {author} {\bibfnamefont {Valentin}\ \bibnamefont {Cr{\'e}pel}}, \bibinfo {author} {\bibfnamefont {Yang}\ \bibnamefont {Zhang}}, \ and\ \bibinfo {author} {\bibfnamefont {Liang}\ \bibnamefont {Fu}},\ }\bibfield  {title} {\enquote {\bibinfo {title} {Magic in twisted transition metal dichalcogenide bilayers},}\ }\href@noop {} {\bibfield  {journal} {\bibinfo  {journal} {Nature communications}\ }\textbf {\bibinfo {volume} {12}},\ \bibinfo {pages} {6730} (\bibinfo {year} {2021}{\natexlab{a}})}\BibitemShut {NoStop}%
\bibitem [{\citenamefont {Cr{\'e}pel}\ \emph {et~al.}(2024)\citenamefont {Cr{\'e}pel}, \citenamefont {Regnault},\ and\ \citenamefont {Queiroz}}]{Crepel2024}%
  \BibitemOpen
  \bibfield  {author} {\bibinfo {author} {\bibfnamefont {Valentin}\ \bibnamefont {Cr{\'e}pel}}, \bibinfo {author} {\bibfnamefont {Nicolas}\ \bibnamefont {Regnault}}, \ and\ \bibinfo {author} {\bibfnamefont {Raquel}\ \bibnamefont {Queiroz}},\ }\bibfield  {title} {\enquote {\bibinfo {title} {Chiral limit and origin of topological flat bands in twisted transition metal dichalcogenide homobilayers},}\ }\href {\doibase 10.1038/s42005-024-01641-6} {\bibfield  {journal} {\bibinfo  {journal} {Communications Physics}\ }\textbf {\bibinfo {volume} {7}},\ \bibinfo {pages} {146} (\bibinfo {year} {2024})}\BibitemShut {NoStop}%
\bibitem [{\citenamefont {Jia}\ \emph {et~al.}(2024)\citenamefont {Jia}, \citenamefont {Yu}, \citenamefont {Liu}, \citenamefont {Herzog-Arbeitman}, \citenamefont {Qi}, \citenamefont {Pi}, \citenamefont {Regnault}, \citenamefont {Weng}, \citenamefont {Bernevig},\ and\ \citenamefont {Wu}}]{WuBAB2024_mote2}%
  \BibitemOpen
  \bibfield  {author} {\bibinfo {author} {\bibfnamefont {Yujin}\ \bibnamefont {Jia}}, \bibinfo {author} {\bibfnamefont {Jiabin}\ \bibnamefont {Yu}}, \bibinfo {author} {\bibfnamefont {Jiaxuan}\ \bibnamefont {Liu}}, \bibinfo {author} {\bibfnamefont {Jonah}\ \bibnamefont {Herzog-Arbeitman}}, \bibinfo {author} {\bibfnamefont {Ziyue}\ \bibnamefont {Qi}}, \bibinfo {author} {\bibfnamefont {Hanqi}\ \bibnamefont {Pi}}, \bibinfo {author} {\bibfnamefont {Nicolas}\ \bibnamefont {Regnault}}, \bibinfo {author} {\bibfnamefont {Hongming}\ \bibnamefont {Weng}}, \bibinfo {author} {\bibfnamefont {B.~Andrei}\ \bibnamefont {Bernevig}}, \ and\ \bibinfo {author} {\bibfnamefont {Quansheng}\ \bibnamefont {Wu}},\ }\bibfield  {title} {\enquote {\bibinfo {title} {Moir\'e fractional chern insulators. i. first-principles calculations and continuum models of twisted bilayer ${\mathrm{mote}}_{2}$},}\ }\href {\doibase 10.1103/PhysRevB.109.205121} {\bibfield  {journal} {\bibinfo  {journal} {Phys. Rev. B}\ }\textbf {\bibinfo {volume} {109}},\
  \bibinfo {pages} {205121} (\bibinfo {year} {2024})}\BibitemShut {NoStop}%
\bibitem [{\citenamefont {{Wang}}\ \emph {et~al.}(2024)\citenamefont {{Wang}}, \citenamefont {{Zhang}}, \citenamefont {{Liu}}, \citenamefont {{Wang}}, \citenamefont {{Cao}},\ and\ \citenamefont {{Xiao}}}]{Xiao2024_mote2}%
  \BibitemOpen
  \bibfield  {author} {\bibinfo {author} {\bibfnamefont {Chong}\ \bibnamefont {{Wang}}}, \bibinfo {author} {\bibfnamefont {Xiao-Wei}\ \bibnamefont {{Zhang}}}, \bibinfo {author} {\bibfnamefont {Xiaoyu}\ \bibnamefont {{Liu}}}, \bibinfo {author} {\bibfnamefont {Jie}\ \bibnamefont {{Wang}}}, \bibinfo {author} {\bibfnamefont {Ting}\ \bibnamefont {{Cao}}}, \ and\ \bibinfo {author} {\bibfnamefont {Di}~\bibnamefont {{Xiao}}},\ }\bibfield  {title} {\enquote {\bibinfo {title} {{Higher Landau-Level Analogues and Signatures of Non-Abelian States in Twisted Bilayer MoTe$_2$}},}\ }\href {\doibase 10.48550/arXiv.2404.05697} {\bibfield  {journal} {\bibinfo  {journal} {arXiv e-prints}\ ,\ \bibinfo {eid} {arXiv:2404.05697}} (\bibinfo {year} {2024})},\ \Eprint {http://arxiv.org/abs/2404.05697} {arXiv:2404.05697 [cond-mat.str-el]} \BibitemShut {NoStop}%
\bibitem [{\citenamefont {{Xu}}\ \emph {et~al.}(2024)\citenamefont {{Xu}}, \citenamefont {{Mao}}, \citenamefont {{Zeng}},\ and\ \citenamefont {{Zhang}}}]{ZhangYang2024_mote2}%
  \BibitemOpen
  \bibfield  {author} {\bibinfo {author} {\bibfnamefont {Cheng}\ \bibnamefont {{Xu}}}, \bibinfo {author} {\bibfnamefont {Ning}\ \bibnamefont {{Mao}}}, \bibinfo {author} {\bibfnamefont {Tiansheng}\ \bibnamefont {{Zeng}}}, \ and\ \bibinfo {author} {\bibfnamefont {Yang}\ \bibnamefont {{Zhang}}},\ }\bibfield  {title} {\enquote {\bibinfo {title} {{Multiple Chern bands in twisted MoTe$_2$ and possible non-Abelian states}},}\ }\href {\doibase 10.48550/arXiv.2403.17003} {\bibfield  {journal} {\bibinfo  {journal} {arXiv e-prints}\ ,\ \bibinfo {eid} {arXiv:2403.17003}} (\bibinfo {year} {2024})},\ \Eprint {http://arxiv.org/abs/2403.17003} {arXiv:2403.17003 [cond-mat.str-el]} \BibitemShut {NoStop}%
\bibitem [{\citenamefont {König}\ \emph {et~al.}(2007)\citenamefont {König}, \citenamefont {Wiedmann}, \citenamefont {Brüne}, \citenamefont {Roth}, \citenamefont {Buhmann}, \citenamefont {Molenkamp}, \citenamefont {Qi},\ and\ \citenamefont {Zhang}}]{QSHex}%
  \BibitemOpen
  \bibfield  {author} {\bibinfo {author} {\bibfnamefont {Markus}\ \bibnamefont {König}}, \bibinfo {author} {\bibfnamefont {Steffen}\ \bibnamefont {Wiedmann}}, \bibinfo {author} {\bibfnamefont {Christoph}\ \bibnamefont {Brüne}}, \bibinfo {author} {\bibfnamefont {Andreas}\ \bibnamefont {Roth}}, \bibinfo {author} {\bibfnamefont {Hartmut}\ \bibnamefont {Buhmann}}, \bibinfo {author} {\bibfnamefont {Laurens~W.}\ \bibnamefont {Molenkamp}}, \bibinfo {author} {\bibfnamefont {Xiao-Liang}\ \bibnamefont {Qi}}, \ and\ \bibinfo {author} {\bibfnamefont {Shou-Cheng}\ \bibnamefont {Zhang}},\ }\bibfield  {title} {\enquote {\bibinfo {title} {Quantum spin hall insulator state in hgte quantum wells},}\ }\href {\doibase 10.1126/science.1148047} {\bibfield  {journal} {\bibinfo  {journal} {Science}\ }\textbf {\bibinfo {volume} {318}},\ \bibinfo {pages} {766--770} (\bibinfo {year} {2007})},\ \Eprint {http://arxiv.org/abs/https://www.science.org/doi/pdf/10.1126/science.1148047}
  {https://www.science.org/doi/pdf/10.1126/science.1148047} \BibitemShut {NoStop}%
\bibitem [{\citenamefont {Lindner}\ \emph {et~al.}(2012)\citenamefont {Lindner}, \citenamefont {Berg}, \citenamefont {Refael},\ and\ \citenamefont {Stern}}]{lidnerstern}%
  \BibitemOpen
  \bibfield  {author} {\bibinfo {author} {\bibfnamefont {Netanel~H.}\ \bibnamefont {Lindner}}, \bibinfo {author} {\bibfnamefont {Erez}\ \bibnamefont {Berg}}, \bibinfo {author} {\bibfnamefont {Gil}\ \bibnamefont {Refael}}, \ and\ \bibinfo {author} {\bibfnamefont {Ady}\ \bibnamefont {Stern}},\ }\bibfield  {title} {\enquote {\bibinfo {title} {Fractionalizing majorana fermions: Non-abelian statistics on the edges of abelian quantum hall states},}\ }\href {\doibase 10.1103/PhysRevX.2.041002} {\bibfield  {journal} {\bibinfo  {journal} {Phys. Rev. X}\ }\textbf {\bibinfo {volume} {2}},\ \bibinfo {pages} {041002} (\bibinfo {year} {2012})}\BibitemShut {NoStop}%
\bibitem [{\citenamefont {Cheng}(2012)}]{cheng2012}%
  \BibitemOpen
  \bibfield  {author} {\bibinfo {author} {\bibfnamefont {Meng}\ \bibnamefont {Cheng}},\ }\bibfield  {title} {\enquote {\bibinfo {title} {Superconducting proximity effect on the edge of fractional topological insulators},}\ }\href {\doibase 10.1103/PhysRevB.86.195126} {\bibfield  {journal} {\bibinfo  {journal} {Phys. Rev. B}\ }\textbf {\bibinfo {volume} {86}},\ \bibinfo {pages} {195126} (\bibinfo {year} {2012})}\BibitemShut {NoStop}%
\bibitem [{\citenamefont {Stern}(2016)}]{Stern_2016}%
  \BibitemOpen
  \bibfield  {author} {\bibinfo {author} {\bibfnamefont {Ady}\ \bibnamefont {Stern}},\ }\bibfield  {title} {\enquote {\bibinfo {title} {Fractional topological insulators: A pedagogical review},}\ }\href {\doibase 10.1146/annurev-conmatphys-031115-011559} {\bibfield  {journal} {\bibinfo  {journal} {Annual Review of Condensed Matter Physics}\ }\textbf {\bibinfo {volume} {7}},\ \bibinfo {pages} {349–368} (\bibinfo {year} {2016})}\BibitemShut {NoStop}%
\bibitem [{\citenamefont {Levin}\ and\ \citenamefont {Stern}(2009)}]{FTI}%
  \BibitemOpen
  \bibfield  {author} {\bibinfo {author} {\bibfnamefont {Michael}\ \bibnamefont {Levin}}\ and\ \bibinfo {author} {\bibfnamefont {Ady}\ \bibnamefont {Stern}},\ }\bibfield  {title} {\enquote {\bibinfo {title} {Fractional topological insulators},}\ }\href {\doibase 10.1103/PhysRevLett.103.196803} {\bibfield  {journal} {\bibinfo  {journal} {Phys. Rev. Lett.}\ }\textbf {\bibinfo {volume} {103}},\ \bibinfo {pages} {196803} (\bibinfo {year} {2009})}\BibitemShut {NoStop}%
\bibitem [{\citenamefont {Levin}\ and\ \citenamefont {Stern}(2012)}]{FTI2}%
  \BibitemOpen
  \bibfield  {author} {\bibinfo {author} {\bibfnamefont {Michael}\ \bibnamefont {Levin}}\ and\ \bibinfo {author} {\bibfnamefont {Ady}\ \bibnamefont {Stern}},\ }\bibfield  {title} {\enquote {\bibinfo {title} {Classification and analysis of two-dimensional abelian fractional topological insulators},}\ }\href {\doibase 10.1103/PhysRevB.86.115131} {\bibfield  {journal} {\bibinfo  {journal} {Phys. Rev. B}\ }\textbf {\bibinfo {volume} {86}},\ \bibinfo {pages} {115131} (\bibinfo {year} {2012})}\BibitemShut {NoStop}%
\bibitem [{\citenamefont {Morales-Dur¨¢n}\ \emph {et~al.}(2023)\citenamefont {Morales-Dur¨¢n}, \citenamefont {Wei},\ and\ \citenamefont {MacDonald}}]{macdonaldLL}%
  \BibitemOpen
  \bibfield  {author} {\bibinfo {author} {\bibfnamefont {Nicol¨¢s}\ \bibnamefont {Morales-Dur¨¢n}}, \bibinfo {author} {\bibfnamefont {Nemin}\ \bibnamefont {Wei}}, \ and\ \bibinfo {author} {\bibfnamefont {Allan~H.}\ \bibnamefont {MacDonald}},\ }\href@noop {} {\enquote {\bibinfo {title} {Magic angles and fractional chern insulators in twisted homobilayer tmds},}\ } (\bibinfo {year} {2023}),\ \Eprint {http://arxiv.org/abs/2308.03143} {arXiv:2308.03143 [cond-mat.str-el]} \BibitemShut {NoStop}%
\bibitem [{\citenamefont {Paul}\ \emph {et~al.}(2023)\citenamefont {Paul}, \citenamefont {Zhang},\ and\ \citenamefont {Fu}}]{Paul_2023}%
  \BibitemOpen
  \bibfield  {author} {\bibinfo {author} {\bibfnamefont {Nisarga}\ \bibnamefont {Paul}}, \bibinfo {author} {\bibfnamefont {Yang}\ \bibnamefont {Zhang}}, \ and\ \bibinfo {author} {\bibfnamefont {Liang}\ \bibnamefont {Fu}},\ }\bibfield  {title} {\enquote {\bibinfo {title} {Giant proximity exchange and flat chern band in 2d magnet-semiconductor heterostructures},}\ }\href {\doibase 10.1126/sciadv.abn1401} {\bibfield  {journal} {\bibinfo  {journal} {Science Advances}\ }\textbf {\bibinfo {volume} {9}} (\bibinfo {year} {2023}),\ 10.1126/sciadv.abn1401}\BibitemShut {NoStop}%
\bibitem [{\citenamefont {Moore}\ and\ \citenamefont {Read}(1991)}]{mooreread}%
  \BibitemOpen
  \bibfield  {author} {\bibinfo {author} {\bibfnamefont {Gregory}\ \bibnamefont {Moore}}\ and\ \bibinfo {author} {\bibfnamefont {Nicholas}\ \bibnamefont {Read}},\ }\bibfield  {title} {\enquote {\bibinfo {title} {Nonabelions in the fractional quantum hall effect},}\ }\href {\doibase https://doi.org/10.1016/0550-3213(91)90407-O} {\bibfield  {journal} {\bibinfo  {journal} {Nuclear Physics B}\ }\textbf {\bibinfo {volume} {360}},\ \bibinfo {pages} {362--396} (\bibinfo {year} {1991})}\BibitemShut {NoStop}%
\bibitem [{\citenamefont {Greiter}\ \emph {et~al.}(1991)\citenamefont {Greiter}, \citenamefont {Wen},\ and\ \citenamefont {Wilczek}}]{pfaffianwen}%
  \BibitemOpen
  \bibfield  {author} {\bibinfo {author} {\bibfnamefont {Martin}\ \bibnamefont {Greiter}}, \bibinfo {author} {\bibfnamefont {Xiao-Gang}\ \bibnamefont {Wen}}, \ and\ \bibinfo {author} {\bibfnamefont {Frank}\ \bibnamefont {Wilczek}},\ }\bibfield  {title} {\enquote {\bibinfo {title} {Paired hall state at half filling},}\ }\href {\doibase 10.1103/PhysRevLett.66.3205} {\bibfield  {journal} {\bibinfo  {journal} {Phys. Rev. Lett.}\ }\textbf {\bibinfo {volume} {66}},\ \bibinfo {pages} {3205--3208} (\bibinfo {year} {1991})}\BibitemShut {NoStop}%
\bibitem [{\citenamefont {Wen}(1993)}]{pfaffianwen2}%
  \BibitemOpen
  \bibfield  {author} {\bibinfo {author} {\bibfnamefont {Xiao-Gang}\ \bibnamefont {Wen}},\ }\bibfield  {title} {\enquote {\bibinfo {title} {Topological order and edge structure of \ensuremath{\nu}=1/2 quantum hall state},}\ }\href {\doibase 10.1103/PhysRevLett.70.355} {\bibfield  {journal} {\bibinfo  {journal} {Phys. Rev. Lett.}\ }\textbf {\bibinfo {volume} {70}},\ \bibinfo {pages} {355--358} (\bibinfo {year} {1993})}\BibitemShut {NoStop}%
\bibitem [{\citenamefont {{Levin}}\ \emph {et~al.}(2007)\citenamefont {{Levin}}, \citenamefont {{Halperin}},\ and\ \citenamefont {{Rosenow}}}]{Levin_AntiPf}%
  \BibitemOpen
  \bibfield  {author} {\bibinfo {author} {\bibfnamefont {Michael}\ \bibnamefont {{Levin}}}, \bibinfo {author} {\bibfnamefont {Bertrand~I.}\ \bibnamefont {{Halperin}}}, \ and\ \bibinfo {author} {\bibfnamefont {Bernd}\ \bibnamefont {{Rosenow}}},\ }\bibfield  {title} {\enquote {\bibinfo {title} {{Particle-Hole Symmetry and the Pfaffian State}},}\ }\href {\doibase 10.1103/PhysRevLett.99.236806} {\bibfield  {journal} {\bibinfo  {journal} {Physical Review Letters}\ }\textbf {\bibinfo {volume} {99}},\ \bibinfo {eid} {236806} (\bibinfo {year} {2007})},\ \Eprint {http://arxiv.org/abs/0707.0483} {arXiv:0707.0483 [cond-mat.mes-hall]} \BibitemShut {NoStop}%
\bibitem [{\citenamefont {Son}(2015)}]{Son2015}%
  \BibitemOpen
  \bibfield  {author} {\bibinfo {author} {\bibfnamefont {Dam~Thanh}\ \bibnamefont {Son}},\ }\bibfield  {title} {\enquote {\bibinfo {title} {Is the composite fermion a dirac particle?}}\ }\href {\doibase 10.1103/PhysRevX.5.031027} {\bibfield  {journal} {\bibinfo  {journal} {Phys. Rev. X}\ }\textbf {\bibinfo {volume} {5}},\ \bibinfo {pages} {031027} (\bibinfo {year} {2015})}\BibitemShut {NoStop}%
\bibitem [{\citenamefont {Greiter}\ \emph {et~al.}(1992)\citenamefont {Greiter}, \citenamefont {Wen},\ and\ \citenamefont {Wilczek}}]{GREITER1992567}%
  \BibitemOpen
  \bibfield  {author} {\bibinfo {author} {\bibfnamefont {Martin}\ \bibnamefont {Greiter}}, \bibinfo {author} {\bibfnamefont {X.G.}\ \bibnamefont {Wen}}, \ and\ \bibinfo {author} {\bibfnamefont {Frank}\ \bibnamefont {Wilczek}},\ }\bibfield  {title} {\enquote {\bibinfo {title} {Paired hall states},}\ }\href {\doibase https://doi.org/10.1016/0550-3213(92)90401-V} {\bibfield  {journal} {\bibinfo  {journal} {Nuclear Physics B}\ }\textbf {\bibinfo {volume} {374}},\ \bibinfo {pages} {567--614} (\bibinfo {year} {1992})}\BibitemShut {NoStop}%
\bibitem [{\citenamefont {Wen}\ and\ \citenamefont {Zee}(1992)}]{WenZee1992}%
  \BibitemOpen
  \bibfield  {author} {\bibinfo {author} {\bibfnamefont {X.~G.}\ \bibnamefont {Wen}}\ and\ \bibinfo {author} {\bibfnamefont {A.}~\bibnamefont {Zee}},\ }\bibfield  {title} {\enquote {\bibinfo {title} {Classification of abelian quantum hall states and matrix formulation of topological fluids},}\ }\href {\doibase 10.1103/PhysRevB.46.2290} {\bibfield  {journal} {\bibinfo  {journal} {Phys. Rev. B}\ }\textbf {\bibinfo {volume} {46}},\ \bibinfo {pages} {2290--2301} (\bibinfo {year} {1992})}\BibitemShut {NoStop}%
\bibitem [{\citenamefont {{Overbosch}}\ and\ \citenamefont {{Wen}}(2008)}]{OverboschWen2008}%
  \BibitemOpen
  \bibfield  {author} {\bibinfo {author} {\bibfnamefont {B.~J.}\ \bibnamefont {{Overbosch}}}\ and\ \bibinfo {author} {\bibfnamefont {Xiao-Gang}\ \bibnamefont {{Wen}}},\ }\bibfield  {title} {\enquote {\bibinfo {title} {{Phase transitions on the edge of the $\nu$=5/2 Pfaffian and anti-Pfaffian quantum Hall state}},}\ }\href {\doibase 10.48550/arXiv.0804.2087} {\bibfield  {journal} {\bibinfo  {journal} {arXiv e-prints}\ ,\ \bibinfo {eid} {arXiv:0804.2087}} (\bibinfo {year} {2008})},\ \Eprint {http://arxiv.org/abs/0804.2087} {arXiv:0804.2087 [cond-mat.mes-hall]} \BibitemShut {NoStop}%
\bibitem [{\citenamefont {{Ahn}}\ \emph {et~al.}(2024)\citenamefont {{Ahn}}, \citenamefont {{Lee}}, \citenamefont {{Yananose}}, \citenamefont {{Kim}},\ and\ \citenamefont {{Cho}}}]{ChoGilYoung_mote2_nonAbelian}%
  \BibitemOpen
  \bibfield  {author} {\bibinfo {author} {\bibfnamefont {Cheong-Eung}\ \bibnamefont {{Ahn}}}, \bibinfo {author} {\bibfnamefont {Wonjun}\ \bibnamefont {{Lee}}}, \bibinfo {author} {\bibfnamefont {Kunihiro}\ \bibnamefont {{Yananose}}}, \bibinfo {author} {\bibfnamefont {Youngwook}\ \bibnamefont {{Kim}}}, \ and\ \bibinfo {author} {\bibfnamefont {Gil~Young}\ \bibnamefont {{Cho}}},\ }\bibfield  {title} {\enquote {\bibinfo {title} {{Non-Abelian Fractional Quantum Anomalous Hall States and First Landau Level Physics in Second Moir{\'e} Band of Twisted Bilayer MoTe2}},}\ }\href {\doibase 10.48550/arXiv.2403.19155} {\bibfield  {journal} {\bibinfo  {journal} {arXiv e-prints}\ ,\ \bibinfo {eid} {arXiv:2403.19155}} (\bibinfo {year} {2024})},\ \Eprint {http://arxiv.org/abs/2403.19155} {arXiv:2403.19155 [cond-mat.str-el]} \BibitemShut {NoStop}%
\bibitem [{\citenamefont {{Reddy}}\ \emph {et~al.}(2024)\citenamefont {{Reddy}}, \citenamefont {{Paul}}, \citenamefont {{Abouelkomsan}},\ and\ \citenamefont {{Fu}}}]{Fu2024NonAbelianTopoMini}%
  \BibitemOpen
  \bibfield  {author} {\bibinfo {author} {\bibfnamefont {Aidan~P.}\ \bibnamefont {{Reddy}}}, \bibinfo {author} {\bibfnamefont {Nisarga}\ \bibnamefont {{Paul}}}, \bibinfo {author} {\bibfnamefont {Ahmed}\ \bibnamefont {{Abouelkomsan}}}, \ and\ \bibinfo {author} {\bibfnamefont {Liang}\ \bibnamefont {{Fu}}},\ }\bibfield  {title} {\enquote {\bibinfo {title} {{Non-Abelian fractionalization in topological minibands}},}\ }\href {\doibase 10.48550/arXiv.2403.00059} {\bibfield  {journal} {\bibinfo  {journal} {arXiv e-prints}\ ,\ \bibinfo {eid} {arXiv:2403.00059}} (\bibinfo {year} {2024})},\ \Eprint {http://arxiv.org/abs/2403.00059} {arXiv:2403.00059 [cond-mat.mes-hall]} \BibitemShut {NoStop}%
\bibitem [{\citenamefont {{Chen}}\ \emph {et~al.}(2024)\citenamefont {{Chen}}, \citenamefont {{Luo}}, \citenamefont {{Zhu}},\ and\ \citenamefont {{Sheng}}}]{Sheng2024nonAbelian}%
  \BibitemOpen
  \bibfield  {author} {\bibinfo {author} {\bibfnamefont {Feng}\ \bibnamefont {{Chen}}}, \bibinfo {author} {\bibfnamefont {Wei-Wei}\ \bibnamefont {{Luo}}}, \bibinfo {author} {\bibfnamefont {Wei}\ \bibnamefont {{Zhu}}}, \ and\ \bibinfo {author} {\bibfnamefont {D.~N.}\ \bibnamefont {{Sheng}}},\ }\bibfield  {title} {\enquote {\bibinfo {title} {{Robust non-Abelian even-denominator fractional Chern insulator in twisted bilayer MoTe$_2$}},}\ }\href {\doibase 10.48550/arXiv.2405.08386} {\bibfield  {journal} {\bibinfo  {journal} {arXiv e-prints}\ ,\ \bibinfo {eid} {arXiv:2405.08386}} (\bibinfo {year} {2024})},\ \Eprint {http://arxiv.org/abs/2405.08386} {arXiv:2405.08386 [cond-mat.str-el]} \BibitemShut {NoStop}%
\bibitem [{\citenamefont {Dong}\ \emph {et~al.}(2023)\citenamefont {Dong}, \citenamefont {Wang}, \citenamefont {Ledwith}, \citenamefont {Vishwanath},\ and\ \citenamefont {Parker}}]{Dong2023CFL}%
  \BibitemOpen
  \bibfield  {author} {\bibinfo {author} {\bibfnamefont {Junkai}\ \bibnamefont {Dong}}, \bibinfo {author} {\bibfnamefont {Jie}\ \bibnamefont {Wang}}, \bibinfo {author} {\bibfnamefont {Patrick~J.}\ \bibnamefont {Ledwith}}, \bibinfo {author} {\bibfnamefont {Ashvin}\ \bibnamefont {Vishwanath}}, \ and\ \bibinfo {author} {\bibfnamefont {Daniel~E.}\ \bibnamefont {Parker}},\ }\bibfield  {title} {\enquote {\bibinfo {title} {Composite fermi liquid at zero magnetic field in twisted ${\mathrm{mote}}_{2}$},}\ }\href {\doibase 10.1103/PhysRevLett.131.136502} {\bibfield  {journal} {\bibinfo  {journal} {Phys. Rev. Lett.}\ }\textbf {\bibinfo {volume} {131}},\ \bibinfo {pages} {136502} (\bibinfo {year} {2023})}\BibitemShut {NoStop}%
\bibitem [{\citenamefont {Goldman}\ \emph {et~al.}(2023)\citenamefont {Goldman}, \citenamefont {Reddy}, \citenamefont {Paul},\ and\ \citenamefont {Fu}}]{Goldman2023CFL}%
  \BibitemOpen
  \bibfield  {author} {\bibinfo {author} {\bibfnamefont {Hart}\ \bibnamefont {Goldman}}, \bibinfo {author} {\bibfnamefont {Aidan~P.}\ \bibnamefont {Reddy}}, \bibinfo {author} {\bibfnamefont {Nisarga}\ \bibnamefont {Paul}}, \ and\ \bibinfo {author} {\bibfnamefont {Liang}\ \bibnamefont {Fu}},\ }\bibfield  {title} {\enquote {\bibinfo {title} {Zero-field composite fermi liquid in twisted semiconductor bilayers},}\ }\href {\doibase 10.1103/PhysRevLett.131.136501} {\bibfield  {journal} {\bibinfo  {journal} {Phys. Rev. Lett.}\ }\textbf {\bibinfo {volume} {131}},\ \bibinfo {pages} {136501} (\bibinfo {year} {2023})}\BibitemShut {NoStop}%
\bibitem [{\citenamefont {Kane}\ and\ \citenamefont {Mele}(2005)}]{kaneqsh}%
  \BibitemOpen
  \bibfield  {author} {\bibinfo {author} {\bibfnamefont {C.~L.}\ \bibnamefont {Kane}}\ and\ \bibinfo {author} {\bibfnamefont {E.~J.}\ \bibnamefont {Mele}},\ }\bibfield  {title} {\enquote {\bibinfo {title} {Quantum spin hall effect in graphene},}\ }\href {\doibase 10.1103/PhysRevLett.95.226801} {\bibfield  {journal} {\bibinfo  {journal} {Phys. Rev. Lett.}\ }\textbf {\bibinfo {volume} {95}},\ \bibinfo {pages} {226801} (\bibinfo {year} {2005})}\BibitemShut {NoStop}%
\bibitem [{\citenamefont {Bernevig}\ and\ \citenamefont {Zhang}(2006)}]{zhangqsh}%
  \BibitemOpen
  \bibfield  {author} {\bibinfo {author} {\bibfnamefont {B.~Andrei}\ \bibnamefont {Bernevig}}\ and\ \bibinfo {author} {\bibfnamefont {Shou-Cheng}\ \bibnamefont {Zhang}},\ }\bibfield  {title} {\enquote {\bibinfo {title} {Quantum spin hall effect},}\ }\href {\doibase 10.1103/PhysRevLett.96.106802} {\bibfield  {journal} {\bibinfo  {journal} {Phys. Rev. Lett.}\ }\textbf {\bibinfo {volume} {96}},\ \bibinfo {pages} {106802} (\bibinfo {year} {2006})}\BibitemShut {NoStop}%
\bibitem [{\citenamefont {{Zhang}}(2024{\natexlab{a}})}]{zhangccfl}%
  \BibitemOpen
  \bibfield  {author} {\bibinfo {author} {\bibfnamefont {Ya-Hui}\ \bibnamefont {{Zhang}}},\ }\bibfield  {title} {\enquote {\bibinfo {title} {{Vortex spin liquid with fractional quantum spin Hall effect in moir{\'e} Chern bands}},}\ }\href {\doibase 10.48550/arXiv.2402.05112} {\bibfield  {journal} {\bibinfo  {journal} {arXiv e-prints}\ ,\ \bibinfo {eid} {arXiv:2402.05112}} (\bibinfo {year} {2024}{\natexlab{a}})},\ \Eprint {http://arxiv.org/abs/2402.05112} {arXiv:2402.05112 [cond-mat.str-el]} \BibitemShut {NoStop}%
\bibitem [{\citenamefont {{Myerson-Jain}}\ \emph {et~al.}(2023)\citenamefont {{Myerson-Jain}}, \citenamefont {{Jian}},\ and\ \citenamefont {{Xu}}}]{nayanccfl}%
  \BibitemOpen
  \bibfield  {author} {\bibinfo {author} {\bibfnamefont {Nayan}\ \bibnamefont {{Myerson-Jain}}}, \bibinfo {author} {\bibfnamefont {Chao-Ming}\ \bibnamefont {{Jian}}}, \ and\ \bibinfo {author} {\bibfnamefont {Cenke}\ \bibnamefont {{Xu}}},\ }\bibfield  {title} {\enquote {\bibinfo {title} {{The Conjugate Composite Fermi Liquid}},}\ }\href {\doibase 10.48550/arXiv.2311.16250} {\bibfield  {journal} {\bibinfo  {journal} {arXiv e-prints}\ ,\ \bibinfo {eid} {arXiv:2311.16250}} (\bibinfo {year} {2023})},\ \Eprint {http://arxiv.org/abs/2311.16250} {arXiv:2311.16250 [cond-mat.str-el]} \BibitemShut {NoStop}%
\bibitem [{\citenamefont {Shi}\ \emph {et~al.}(2024)\citenamefont {Shi}, \citenamefont {Goldman}, \citenamefont {Dong},\ and\ \citenamefont {Senthil}}]{shi2024excitonic}%
  \BibitemOpen
  \bibfield  {author} {\bibinfo {author} {\bibfnamefont {Zhengyan~Darius}\ \bibnamefont {Shi}}, \bibinfo {author} {\bibfnamefont {Hart}\ \bibnamefont {Goldman}}, \bibinfo {author} {\bibfnamefont {Zhihuan}\ \bibnamefont {Dong}}, \ and\ \bibinfo {author} {\bibfnamefont {T.}~\bibnamefont {Senthil}},\ }\href@noop {} {\enquote {\bibinfo {title} {Excitonic quantum criticality: from bilayer graphene to narrow chern bands},}\ } (\bibinfo {year} {2024}),\ \Eprint {http://arxiv.org/abs/2402.12436} {arXiv:2402.12436 [cond-mat.str-el]} \BibitemShut {NoStop}%
\bibitem [{\citenamefont {{Kitaev}}(2003)}]{Kitaev2003}%
  \BibitemOpen
  \bibfield  {author} {\bibinfo {author} {\bibfnamefont {A.~Yu.}\ \bibnamefont {{Kitaev}}},\ }\bibfield  {title} {\enquote {\bibinfo {title} {{Fault-tolerant quantum computation by anyons}},}\ }\href {\doibase 10.1016/S0003-4916(02)00018-0} {\bibfield  {journal} {\bibinfo  {journal} {Annals of Physics}\ }\textbf {\bibinfo {volume} {303}},\ \bibinfo {pages} {2--30} (\bibinfo {year} {2003})},\ \Eprint {http://arxiv.org/abs/quant-ph/9707021} {arXiv:quant-ph/9707021 [quant-ph]} \BibitemShut {NoStop}%
\bibitem [{\citenamefont {Wen}(1990)}]{wenedge1}%
  \BibitemOpen
  \bibfield  {author} {\bibinfo {author} {\bibfnamefont {X.~G.}\ \bibnamefont {Wen}},\ }\bibfield  {title} {\enquote {\bibinfo {title} {Electrodynamical properties of gapless edge excitations in the fractional quantum hall states},}\ }\href {\doibase 10.1103/PhysRevLett.64.2206} {\bibfield  {journal} {\bibinfo  {journal} {Phys. Rev. Lett.}\ }\textbf {\bibinfo {volume} {64}},\ \bibinfo {pages} {2206--2209} (\bibinfo {year} {1990})}\BibitemShut {NoStop}%
\bibitem [{\citenamefont {Wen}(1991)}]{wenedge2}%
  \BibitemOpen
  \bibfield  {author} {\bibinfo {author} {\bibfnamefont {X.~G.}\ \bibnamefont {Wen}},\ }\bibfield  {title} {\enquote {\bibinfo {title} {Gapless boundary excitations in the quantum hall states and in the chiral spin states},}\ }\href {\doibase 10.1103/PhysRevB.43.11025} {\bibfield  {journal} {\bibinfo  {journal} {Phys. Rev. B}\ }\textbf {\bibinfo {volume} {43}},\ \bibinfo {pages} {11025--11036} (\bibinfo {year} {1991})}\BibitemShut {NoStop}%
\bibitem [{Note1()}]{Note1}%
  \BibitemOpen
  \bibinfo {note} {In principle, one can consider a more general time-reversal action with $\phi ^c \rightarrow -\phi ^c + \alpha _0$ for some finite constant $\alpha _0$. However, this constant $\alpha _0$ can be absorbed by redefining the time-reversal symmetry as the original one followed by an extra charge U(1) rotation. The redefined time-reversal action remains to be a symmetry action of order 2. We caution that one cannot redefine the time-reversal symmetry by combining it with arbitrary spin $S_z$ rotations because the resulting action is generically not an order-2 action anymore.}\BibitemShut {Stop}%
\bibitem [{\citenamefont {Kane}\ \emph {et~al.}(1994)\citenamefont {Kane}, \citenamefont {Fisher},\ and\ \citenamefont {Polchinski}}]{kanefisherpol}%
  \BibitemOpen
  \bibfield  {author} {\bibinfo {author} {\bibfnamefont {C.~L.}\ \bibnamefont {Kane}}, \bibinfo {author} {\bibfnamefont {Matthew P.~A.}\ \bibnamefont {Fisher}}, \ and\ \bibinfo {author} {\bibfnamefont {J.}~\bibnamefont {Polchinski}},\ }\bibfield  {title} {\enquote {\bibinfo {title} {Randomness at the edge: Theory of quantum hall transport at filling \ensuremath{\nu}=2/3},}\ }\href {\doibase 10.1103/PhysRevLett.72.4129} {\bibfield  {journal} {\bibinfo  {journal} {Phys. Rev. Lett.}\ }\textbf {\bibinfo {volume} {72}},\ \bibinfo {pages} {4129--4132} (\bibinfo {year} {1994})}\BibitemShut {NoStop}%
\bibitem [{\citenamefont {Kane}\ and\ \citenamefont {Fisher}(1995{\natexlab{a}})}]{kanefisher1}%
  \BibitemOpen
  \bibfield  {author} {\bibinfo {author} {\bibfnamefont {C.~L.}\ \bibnamefont {Kane}}\ and\ \bibinfo {author} {\bibfnamefont {Matthew P.~A.}\ \bibnamefont {Fisher}},\ }\bibfield  {title} {\enquote {\bibinfo {title} {Impurity scattering and transport of fractional quantum hall edge states},}\ }\href {\doibase 10.1103/PhysRevB.51.13449} {\bibfield  {journal} {\bibinfo  {journal} {Phys. Rev. B}\ }\textbf {\bibinfo {volume} {51}},\ \bibinfo {pages} {13449--13466} (\bibinfo {year} {1995}{\natexlab{a}})}\BibitemShut {NoStop}%
\bibitem [{\citenamefont {Kane}\ and\ \citenamefont {Fisher}(1995{\natexlab{b}})}]{kanefisher2}%
  \BibitemOpen
  \bibfield  {author} {\bibinfo {author} {\bibfnamefont {C.~L.}\ \bibnamefont {Kane}}\ and\ \bibinfo {author} {\bibfnamefont {Matthew P.~A.}\ \bibnamefont {Fisher}},\ }\bibfield  {title} {\enquote {\bibinfo {title} {Contacts and edge-state equilibration in the fractional quantum hall effect},}\ }\href {\doibase 10.1103/PhysRevB.52.17393} {\bibfield  {journal} {\bibinfo  {journal} {Phys. Rev. B}\ }\textbf {\bibinfo {volume} {52}},\ \bibinfo {pages} {17393--17405} (\bibinfo {year} {1995}{\natexlab{b}})}\BibitemShut {NoStop}%
\bibitem [{\citenamefont {Xu}\ and\ \citenamefont {Moore}(2006)}]{xuedge}%
  \BibitemOpen
  \bibfield  {author} {\bibinfo {author} {\bibfnamefont {Cenke}\ \bibnamefont {Xu}}\ and\ \bibinfo {author} {\bibfnamefont {J.~E.}\ \bibnamefont {Moore}},\ }\bibfield  {title} {\enquote {\bibinfo {title} {Stability of the quantum spin hall effect: Effects of interactions, disorder, and ${\mathbb{z}}_{2}$ topology},}\ }\href {\doibase 10.1103/PhysRevB.73.045322} {\bibfield  {journal} {\bibinfo  {journal} {Phys. Rev. B}\ }\textbf {\bibinfo {volume} {73}},\ \bibinfo {pages} {045322} (\bibinfo {year} {2006})}\BibitemShut {NoStop}%
\bibitem [{\citenamefont {Wu}\ \emph {et~al.}(2006)\citenamefont {Wu}, \citenamefont {Bernevig},\ and\ \citenamefont {Zhang}}]{wuedge}%
  \BibitemOpen
  \bibfield  {author} {\bibinfo {author} {\bibfnamefont {Congjun}\ \bibnamefont {Wu}}, \bibinfo {author} {\bibfnamefont {B.~Andrei}\ \bibnamefont {Bernevig}}, \ and\ \bibinfo {author} {\bibfnamefont {Shou-Cheng}\ \bibnamefont {Zhang}},\ }\bibfield  {title} {\enquote {\bibinfo {title} {Helical liquid and the edge of quantum spin hall systems},}\ }\href {\doibase 10.1103/PhysRevLett.96.106401} {\bibfield  {journal} {\bibinfo  {journal} {Phys. Rev. Lett.}\ }\textbf {\bibinfo {volume} {96}},\ \bibinfo {pages} {106401} (\bibinfo {year} {2006})}\BibitemShut {NoStop}%
\bibitem [{\citenamefont {Bi}\ \emph {et~al.}(2017)\citenamefont {Bi}, \citenamefont {Zhang}, \citenamefont {You}, \citenamefont {Young}, \citenamefont {Balents}, \citenamefont {Liu},\ and\ \citenamefont {Xu}}]{xubsptgraphene}%
  \BibitemOpen
  \bibfield  {author} {\bibinfo {author} {\bibfnamefont {Zhen}\ \bibnamefont {Bi}}, \bibinfo {author} {\bibfnamefont {Ruixing}\ \bibnamefont {Zhang}}, \bibinfo {author} {\bibfnamefont {Yi-Zhuang}\ \bibnamefont {You}}, \bibinfo {author} {\bibfnamefont {Andrea}\ \bibnamefont {Young}}, \bibinfo {author} {\bibfnamefont {Leon}\ \bibnamefont {Balents}}, \bibinfo {author} {\bibfnamefont {Chao-Xing}\ \bibnamefont {Liu}}, \ and\ \bibinfo {author} {\bibfnamefont {Cenke}\ \bibnamefont {Xu}},\ }\bibfield  {title} {\enquote {\bibinfo {title} {Bilayer graphene as a platform for bosonic symmetry-protected topological states},}\ }\href {\doibase 10.1103/PhysRevLett.118.126801} {\bibfield  {journal} {\bibinfo  {journal} {Phys. Rev. Lett.}\ }\textbf {\bibinfo {volume} {118}},\ \bibinfo {pages} {126801} (\bibinfo {year} {2017})}\BibitemShut {NoStop}%
\bibitem [{\citenamefont {Saminadayar}\ \emph {et~al.}(1997)\citenamefont {Saminadayar}, \citenamefont {Glattli}, \citenamefont {Jin},\ and\ \citenamefont {Etienne}}]{Saminadayar_ShotNoise}%
  \BibitemOpen
  \bibfield  {author} {\bibinfo {author} {\bibfnamefont {L.}~\bibnamefont {Saminadayar}}, \bibinfo {author} {\bibfnamefont {D.~C.}\ \bibnamefont {Glattli}}, \bibinfo {author} {\bibfnamefont {Y.}~\bibnamefont {Jin}}, \ and\ \bibinfo {author} {\bibfnamefont {B.}~\bibnamefont {Etienne}},\ }\bibfield  {title} {\enquote {\bibinfo {title} {Observation of the $\mathit{e}\mathit{/}3$ fractionally charged laughlin quasiparticle},}\ }\href {\doibase 10.1103/PhysRevLett.79.2526} {\bibfield  {journal} {\bibinfo  {journal} {Phys. Rev. Lett.}\ }\textbf {\bibinfo {volume} {79}},\ \bibinfo {pages} {2526--2529} (\bibinfo {year} {1997})}\BibitemShut {NoStop}%
\bibitem [{\citenamefont {Martin}\ \emph {et~al.}(2004)\citenamefont {Martin}, \citenamefont {Ilani}, \citenamefont {Verdene}, \citenamefont {Smet}, \citenamefont {Umansky}, \citenamefont {Mahalu}, \citenamefont {Schuh}, \citenamefont {Abstreiter},\ and\ \citenamefont {Yacoby}}]{Yacoby_FractionalCharge_SET}%
  \BibitemOpen
  \bibfield  {author} {\bibinfo {author} {\bibfnamefont {Jens}\ \bibnamefont {Martin}}, \bibinfo {author} {\bibfnamefont {Shahal}\ \bibnamefont {Ilani}}, \bibinfo {author} {\bibfnamefont {Basile}\ \bibnamefont {Verdene}}, \bibinfo {author} {\bibfnamefont {Jurgen}\ \bibnamefont {Smet}}, \bibinfo {author} {\bibfnamefont {Vladimir}\ \bibnamefont {Umansky}}, \bibinfo {author} {\bibfnamefont {Diana}\ \bibnamefont {Mahalu}}, \bibinfo {author} {\bibfnamefont {Dieter}\ \bibnamefont {Schuh}}, \bibinfo {author} {\bibfnamefont {Gerhard}\ \bibnamefont {Abstreiter}}, \ and\ \bibinfo {author} {\bibfnamefont {Amir}\ \bibnamefont {Yacoby}},\ }\bibfield  {title} {\enquote {\bibinfo {title} {Localization of fractionally charged quasi-particles},}\ }\href {\doibase 10.1126/science.1099950} {\bibfield  {journal} {\bibinfo  {journal} {Science}\ }\textbf {\bibinfo {volume} {305}},\ \bibinfo {pages} {980--983} (\bibinfo {year} {2004})},\ \Eprint {http://arxiv.org/abs/https://www.science.org/doi/pdf/10.1126/science.1099950}
  {https://www.science.org/doi/pdf/10.1126/science.1099950} \BibitemShut {NoStop}%
\bibitem [{\citenamefont {Kou}\ \emph {et~al.}(2012)\citenamefont {Kou}, \citenamefont {Marcus}, \citenamefont {Pfeiffer},\ and\ \citenamefont {West}}]{Kou_Antidots}%
  \BibitemOpen
  \bibfield  {author} {\bibinfo {author} {\bibfnamefont {A.}~\bibnamefont {Kou}}, \bibinfo {author} {\bibfnamefont {C.~M.}\ \bibnamefont {Marcus}}, \bibinfo {author} {\bibfnamefont {L.~N.}\ \bibnamefont {Pfeiffer}}, \ and\ \bibinfo {author} {\bibfnamefont {K.~W.}\ \bibnamefont {West}},\ }\bibfield  {title} {\enquote {\bibinfo {title} {Coulomb oscillations in antidots in the integer and fractional quantum hall regimes},}\ }\href {\doibase 10.1103/PhysRevLett.108.256803} {\bibfield  {journal} {\bibinfo  {journal} {Phys. Rev. Lett.}\ }\textbf {\bibinfo {volume} {108}},\ \bibinfo {pages} {256803} (\bibinfo {year} {2012})}\BibitemShut {NoStop}%
\bibitem [{\citenamefont {Kivelson}(1990)}]{KevilsonInterfero}%
  \BibitemOpen
  \bibfield  {author} {\bibinfo {author} {\bibfnamefont {Steven}\ \bibnamefont {Kivelson}},\ }\bibfield  {title} {\enquote {\bibinfo {title} {Semiclassical theory of localized many-anyon states},}\ }\href {\doibase 10.1103/PhysRevLett.65.3369} {\bibfield  {journal} {\bibinfo  {journal} {Phys. Rev. Lett.}\ }\textbf {\bibinfo {volume} {65}},\ \bibinfo {pages} {3369--3372} (\bibinfo {year} {1990})}\BibitemShut {NoStop}%
\bibitem [{\citenamefont {de~C.~Chamon}\ \emph {et~al.}(1997)\citenamefont {de~C.~Chamon}, \citenamefont {Freed}, \citenamefont {Kivelson}, \citenamefont {Sondhi},\ and\ \citenamefont {Wen}}]{WenHallInterfero}%
  \BibitemOpen
  \bibfield  {author} {\bibinfo {author} {\bibfnamefont {C.}~\bibnamefont {de~C.~Chamon}}, \bibinfo {author} {\bibfnamefont {D.~E.}\ \bibnamefont {Freed}}, \bibinfo {author} {\bibfnamefont {S.~A.}\ \bibnamefont {Kivelson}}, \bibinfo {author} {\bibfnamefont {S.~L.}\ \bibnamefont {Sondhi}}, \ and\ \bibinfo {author} {\bibfnamefont {X.~G.}\ \bibnamefont {Wen}},\ }\bibfield  {title} {\enquote {\bibinfo {title} {Two point-contact interferometer for quantum hall systems},}\ }\href {\doibase 10.1103/PhysRevB.55.2331} {\bibfield  {journal} {\bibinfo  {journal} {Phys. Rev. B}\ }\textbf {\bibinfo {volume} {55}},\ \bibinfo {pages} {2331--2343} (\bibinfo {year} {1997})}\BibitemShut {NoStop}%
\bibitem [{\citenamefont {Halperin}\ \emph {et~al.}(2011)\citenamefont {Halperin}, \citenamefont {Stern}, \citenamefont {Neder},\ and\ \citenamefont {Rosenow}}]{HalperinInterfero}%
  \BibitemOpen
  \bibfield  {author} {\bibinfo {author} {\bibfnamefont {Bertrand~I.}\ \bibnamefont {Halperin}}, \bibinfo {author} {\bibfnamefont {Ady}\ \bibnamefont {Stern}}, \bibinfo {author} {\bibfnamefont {Izhar}\ \bibnamefont {Neder}}, \ and\ \bibinfo {author} {\bibfnamefont {Bernd}\ \bibnamefont {Rosenow}},\ }\bibfield  {title} {\enquote {\bibinfo {title} {Theory of the fabry-p\'erot quantum hall interferometer},}\ }\href {\doibase 10.1103/PhysRevB.83.155440} {\bibfield  {journal} {\bibinfo  {journal} {Phys. Rev. B}\ }\textbf {\bibinfo {volume} {83}},\ \bibinfo {pages} {155440} (\bibinfo {year} {2011})}\BibitemShut {NoStop}%
\bibitem [{\citenamefont {{Nakamura}}\ \emph {et~al.}(2019)\citenamefont {{Nakamura}}, \citenamefont {{Fallahi}}, \citenamefont {{Sahasrabudhe}}, \citenamefont {{Rahman}}, \citenamefont {{Liang}}, \citenamefont {{Gardner}},\ and\ \citenamefont {{Manfra}}}]{NakamuraNatPhys2019}%
  \BibitemOpen
  \bibfield  {author} {\bibinfo {author} {\bibfnamefont {J.}~\bibnamefont {{Nakamura}}}, \bibinfo {author} {\bibfnamefont {S.}~\bibnamefont {{Fallahi}}}, \bibinfo {author} {\bibfnamefont {H.}~\bibnamefont {{Sahasrabudhe}}}, \bibinfo {author} {\bibfnamefont {R.}~\bibnamefont {{Rahman}}}, \bibinfo {author} {\bibfnamefont {S.}~\bibnamefont {{Liang}}}, \bibinfo {author} {\bibfnamefont {G.~C.}\ \bibnamefont {{Gardner}}}, \ and\ \bibinfo {author} {\bibfnamefont {M.~J.}\ \bibnamefont {{Manfra}}},\ }\bibfield  {title} {\enquote {\bibinfo {title} {{Aharonov{\textendash}Bohm interference of fractional quantum Hall edge modes}},}\ }\href {\doibase 10.1038/s41567-019-0441-8} {\bibfield  {journal} {\bibinfo  {journal} {Nature Physics}\ }\textbf {\bibinfo {volume} {15}},\ \bibinfo {pages} {563--569} (\bibinfo {year} {2019})},\ \Eprint {http://arxiv.org/abs/1901.08452} {arXiv:1901.08452 [cond-mat.mes-hall]} \BibitemShut {NoStop}%
\bibitem [{\citenamefont {{Nakamura}}\ \emph {et~al.}(2020)\citenamefont {{Nakamura}}, \citenamefont {{Liang}}, \citenamefont {{Gardner}},\ and\ \citenamefont {{Manfra}}}]{NakamuraNatPhys2020}%
  \BibitemOpen
  \bibfield  {author} {\bibinfo {author} {\bibfnamefont {J.}~\bibnamefont {{Nakamura}}}, \bibinfo {author} {\bibfnamefont {S.}~\bibnamefont {{Liang}}}, \bibinfo {author} {\bibfnamefont {G.~C.}\ \bibnamefont {{Gardner}}}, \ and\ \bibinfo {author} {\bibfnamefont {M.~J.}\ \bibnamefont {{Manfra}}},\ }\bibfield  {title} {\enquote {\bibinfo {title} {{Direct observation of anyonic braiding statistics}},}\ }\href {\doibase 10.1038/s41567-020-1019-1} {\bibfield  {journal} {\bibinfo  {journal} {Nature Physics}\ }\textbf {\bibinfo {volume} {16}},\ \bibinfo {pages} {931--936} (\bibinfo {year} {2020})},\ \Eprint {http://arxiv.org/abs/2006.14115} {arXiv:2006.14115 [cond-mat.mes-hall]} \BibitemShut {NoStop}%
\bibitem [{\citenamefont {Nakamura}\ \emph {et~al.}(2023)\citenamefont {Nakamura}, \citenamefont {Liang}, \citenamefont {Gardner},\ and\ \citenamefont {Manfra}}]{NakamuraPRX2023}%
  \BibitemOpen
  \bibfield  {author} {\bibinfo {author} {\bibfnamefont {J.}~\bibnamefont {Nakamura}}, \bibinfo {author} {\bibfnamefont {S.}~\bibnamefont {Liang}}, \bibinfo {author} {\bibfnamefont {G.~C.}\ \bibnamefont {Gardner}}, \ and\ \bibinfo {author} {\bibfnamefont {M.~J.}\ \bibnamefont {Manfra}},\ }\bibfield  {title} {\enquote {\bibinfo {title} {Fabry-p\'erot interferometry at the $\ensuremath{\nu}=2/5$ fractional quantum hall state},}\ }\href {\doibase 10.1103/PhysRevX.13.041012} {\bibfield  {journal} {\bibinfo  {journal} {Phys. Rev. X}\ }\textbf {\bibinfo {volume} {13}},\ \bibinfo {pages} {041012} (\bibinfo {year} {2023})}\BibitemShut {NoStop}%
\bibitem [{\citenamefont {Willett}\ \emph {et~al.}(2023)\citenamefont {Willett}, \citenamefont {Shtengel}, \citenamefont {Nayak}, \citenamefont {Pfeiffer}, \citenamefont {Chung}, \citenamefont {Peabody}, \citenamefont {Baldwin},\ and\ \citenamefont {West}}]{West2023PRX}%
  \BibitemOpen
  \bibfield  {author} {\bibinfo {author} {\bibfnamefont {R.~L.}\ \bibnamefont {Willett}}, \bibinfo {author} {\bibfnamefont {K.}~\bibnamefont {Shtengel}}, \bibinfo {author} {\bibfnamefont {C.}~\bibnamefont {Nayak}}, \bibinfo {author} {\bibfnamefont {L.~N.}\ \bibnamefont {Pfeiffer}}, \bibinfo {author} {\bibfnamefont {Y.~J.}\ \bibnamefont {Chung}}, \bibinfo {author} {\bibfnamefont {M.~L.}\ \bibnamefont {Peabody}}, \bibinfo {author} {\bibfnamefont {K.~W.}\ \bibnamefont {Baldwin}}, \ and\ \bibinfo {author} {\bibfnamefont {K.~W.}\ \bibnamefont {West}},\ }\bibfield  {title} {\enquote {\bibinfo {title} {Interference measurements of non-abelian $e/4$ and abelian $e/2$ quasiparticle braiding},}\ }\href {\doibase 10.1103/PhysRevX.13.011028} {\bibfield  {journal} {\bibinfo  {journal} {Phys. Rev. X}\ }\textbf {\bibinfo {volume} {13}},\ \bibinfo {pages} {011028} (\bibinfo {year} {2023})}\BibitemShut {NoStop}%
\bibitem [{\citenamefont {McClure}\ \emph {et~al.}(2009)\citenamefont {McClure}, \citenamefont {Zhang}, \citenamefont {Rosenow}, \citenamefont {Levenson-Falk}, \citenamefont {Marcus}, \citenamefont {Pfeiffer},\ and\ \citenamefont {West}}]{West2009PRL}%
  \BibitemOpen
  \bibfield  {author} {\bibinfo {author} {\bibfnamefont {D.~T.}\ \bibnamefont {McClure}}, \bibinfo {author} {\bibfnamefont {Yiming}\ \bibnamefont {Zhang}}, \bibinfo {author} {\bibfnamefont {B.}~\bibnamefont {Rosenow}}, \bibinfo {author} {\bibfnamefont {E.~M.}\ \bibnamefont {Levenson-Falk}}, \bibinfo {author} {\bibfnamefont {C.~M.}\ \bibnamefont {Marcus}}, \bibinfo {author} {\bibfnamefont {L.~N.}\ \bibnamefont {Pfeiffer}}, \ and\ \bibinfo {author} {\bibfnamefont {K.~W.}\ \bibnamefont {West}},\ }\bibfield  {title} {\enquote {\bibinfo {title} {Edge-state velocity and coherence in a quantum hall fabry-p\'erot interferometer},}\ }\href {\doibase 10.1103/PhysRevLett.103.206806} {\bibfield  {journal} {\bibinfo  {journal} {Phys. Rev. Lett.}\ }\textbf {\bibinfo {volume} {103}},\ \bibinfo {pages} {206806} (\bibinfo {year} {2009})}\BibitemShut {NoStop}%
\bibitem [{\citenamefont {Camino}\ \emph {et~al.}(2005)\citenamefont {Camino}, \citenamefont {Zhou},\ and\ \citenamefont {Goldman}}]{Camino2005PRB}%
  \BibitemOpen
  \bibfield  {author} {\bibinfo {author} {\bibfnamefont {F.~E.}\ \bibnamefont {Camino}}, \bibinfo {author} {\bibfnamefont {Wei}\ \bibnamefont {Zhou}}, \ and\ \bibinfo {author} {\bibfnamefont {V.~J.}\ \bibnamefont {Goldman}},\ }\bibfield  {title} {\enquote {\bibinfo {title} {Realization of a laughlin quasiparticle interferometer: Observation of fractional statistics},}\ }\href {\doibase 10.1103/PhysRevB.72.075342} {\bibfield  {journal} {\bibinfo  {journal} {Phys. Rev. B}\ }\textbf {\bibinfo {volume} {72}},\ \bibinfo {pages} {075342} (\bibinfo {year} {2005})}\BibitemShut {NoStop}%
\bibitem [{\citenamefont {Camino}\ \emph {et~al.}(2007)\citenamefont {Camino}, \citenamefont {Zhou},\ and\ \citenamefont {Goldman}}]{Camino2007PRL}%
  \BibitemOpen
  \bibfield  {author} {\bibinfo {author} {\bibfnamefont {F.~E.}\ \bibnamefont {Camino}}, \bibinfo {author} {\bibfnamefont {Wei}\ \bibnamefont {Zhou}}, \ and\ \bibinfo {author} {\bibfnamefont {V.~J.}\ \bibnamefont {Goldman}},\ }\bibfield  {title} {\enquote {\bibinfo {title} {$e/3$ laughlin quasiparticle primary-filling $\ensuremath{\nu}=1/3$ interferometer},}\ }\href {\doibase 10.1103/PhysRevLett.98.076805} {\bibfield  {journal} {\bibinfo  {journal} {Phys. Rev. Lett.}\ }\textbf {\bibinfo {volume} {98}},\ \bibinfo {pages} {076805} (\bibinfo {year} {2007})}\BibitemShut {NoStop}%
\bibitem [{\citenamefont {Ofek}\ \emph {et~al.}(2010)\citenamefont {Ofek}, \citenamefont {Bid}, \citenamefont {Heiblum}, \citenamefont {Stern}, \citenamefont {Umansky},\ and\ \citenamefont {Mahalu}}]{Mahalu2010PNAS}%
  \BibitemOpen
  \bibfield  {author} {\bibinfo {author} {\bibfnamefont {Nissim}\ \bibnamefont {Ofek}}, \bibinfo {author} {\bibfnamefont {Aveek}\ \bibnamefont {Bid}}, \bibinfo {author} {\bibfnamefont {Moty}\ \bibnamefont {Heiblum}}, \bibinfo {author} {\bibfnamefont {Ady}\ \bibnamefont {Stern}}, \bibinfo {author} {\bibfnamefont {Vladimir}\ \bibnamefont {Umansky}}, \ and\ \bibinfo {author} {\bibfnamefont {Diana}\ \bibnamefont {Mahalu}},\ }\bibfield  {title} {\enquote {\bibinfo {title} {Role of interactions in an electronic fabry–perot interferometer operating in the quantum hall effect regime},}\ }\href {\doibase 10.1073/pnas.0912624107} {\bibfield  {journal} {\bibinfo  {journal} {Proceedings of the National Academy of Sciences}\ }\textbf {\bibinfo {volume} {107}},\ \bibinfo {pages} {5276--5281} (\bibinfo {year} {2010})},\ \Eprint {http://arxiv.org/abs/https://www.pnas.org/doi/pdf/10.1073/pnas.0912624107} {https://www.pnas.org/doi/pdf/10.1073/pnas.0912624107} \BibitemShut {NoStop}%
\bibitem [{\citenamefont {{Samuelson}}\ \emph {et~al.}(2024)\citenamefont {{Samuelson}}, \citenamefont {{Cohen}}, \citenamefont {{Wang}}, \citenamefont {{Blanch}}, \citenamefont {{Taniguchi}}, \citenamefont {{Watanabe}}, \citenamefont {{Zaletel}},\ and\ \citenamefont {{Young}}}]{Young2024}%
  \BibitemOpen
  \bibfield  {author} {\bibinfo {author} {\bibfnamefont {Noah~L.}\ \bibnamefont {{Samuelson}}}, \bibinfo {author} {\bibfnamefont {Liam~A.}\ \bibnamefont {{Cohen}}}, \bibinfo {author} {\bibfnamefont {Will}\ \bibnamefont {{Wang}}}, \bibinfo {author} {\bibfnamefont {Simon}\ \bibnamefont {{Blanch}}}, \bibinfo {author} {\bibfnamefont {Takashi}\ \bibnamefont {{Taniguchi}}}, \bibinfo {author} {\bibfnamefont {Kenji}\ \bibnamefont {{Watanabe}}}, \bibinfo {author} {\bibfnamefont {Michael~P.}\ \bibnamefont {{Zaletel}}}, \ and\ \bibinfo {author} {\bibfnamefont {Andrea~F.}\ \bibnamefont {{Young}}},\ }\bibfield  {title} {\enquote {\bibinfo {title} {{Anyonic statistics and slow quasiparticle dynamics in a graphene fractional quantum Hall interferometer}},}\ }\href {\doibase 10.48550/arXiv.2403.19628} {\bibfield  {journal} {\bibinfo  {journal} {arXiv e-prints}\ ,\ \bibinfo {eid} {arXiv:2403.19628}} (\bibinfo {year} {2024})},\ \Eprint {http://arxiv.org/abs/2403.19628} {arXiv:2403.19628 [cond-mat.mes-hall]} \BibitemShut
  {NoStop}%
\bibitem [{\citenamefont {{Bonderson}}\ \emph {et~al.}(2008)\citenamefont {{Bonderson}}, \citenamefont {{Shtengel}},\ and\ \citenamefont {{Slingerland}}}]{Bonderson2008Interferometry}%
  \BibitemOpen
  \bibfield  {author} {\bibinfo {author} {\bibfnamefont {Parsa}\ \bibnamefont {{Bonderson}}}, \bibinfo {author} {\bibfnamefont {Kirill}\ \bibnamefont {{Shtengel}}}, \ and\ \bibinfo {author} {\bibfnamefont {J.~K.}\ \bibnamefont {{Slingerland}}},\ }\bibfield  {title} {\enquote {\bibinfo {title} {{Interferometry of non-Abelian anyons}},}\ }\href {\doibase 10.1016/j.aop.2008.01.012} {\bibfield  {journal} {\bibinfo  {journal} {Annals of Physics}\ }\textbf {\bibinfo {volume} {323}},\ \bibinfo {pages} {2709--2755} (\bibinfo {year} {2008})},\ \Eprint {http://arxiv.org/abs/0707.4206} {arXiv:0707.4206 [quant-ph]} \BibitemShut {NoStop}%
\bibitem [{\citenamefont {Cheng}\ \emph {et~al.}(2016)\citenamefont {Cheng}, \citenamefont {Zaletel}, \citenamefont {Barkeshli}, \citenamefont {Vishwanath},\ and\ \citenamefont {Bonderson}}]{Cheng_translation}%
  \BibitemOpen
  \bibfield  {author} {\bibinfo {author} {\bibfnamefont {Meng}\ \bibnamefont {Cheng}}, \bibinfo {author} {\bibfnamefont {Michael}\ \bibnamefont {Zaletel}}, \bibinfo {author} {\bibfnamefont {Maissam}\ \bibnamefont {Barkeshli}}, \bibinfo {author} {\bibfnamefont {Ashvin}\ \bibnamefont {Vishwanath}}, \ and\ \bibinfo {author} {\bibfnamefont {Parsa}\ \bibnamefont {Bonderson}},\ }\bibfield  {title} {\enquote {\bibinfo {title} {Translational symmetry and microscopic constraints on symmetry-enriched topological phases: A view from the surface},}\ }\href {\doibase 10.1103/PhysRevX.6.041068} {\bibfield  {journal} {\bibinfo  {journal} {Phys. Rev. X}\ }\textbf {\bibinfo {volume} {6}},\ \bibinfo {pages} {041068} (\bibinfo {year} {2016})}\BibitemShut {NoStop}%
\bibitem [{\citenamefont {{Kitaev}}(2006)}]{kitaev_2006}%
  \BibitemOpen
  \bibfield  {author} {\bibinfo {author} {\bibfnamefont {Alexei}\ \bibnamefont {{Kitaev}}},\ }\bibfield  {title} {\enquote {\bibinfo {title} {{Anyons in an exactly solved model and beyond}},}\ }\href {\doibase 10.1016/j.aop.2005.10.005} {\bibfield  {journal} {\bibinfo  {journal} {Annals of Physics}\ }\textbf {\bibinfo {volume} {321}},\ \bibinfo {pages} {2--111} (\bibinfo {year} {2006})},\ \Eprint {http://arxiv.org/abs/cond-mat/0506438} {arXiv:cond-mat/0506438 [cond-mat.mes-hall]} \BibitemShut {NoStop}%
\bibitem [{\citenamefont {{Kitaev}}\ and\ \citenamefont {{Preskill}}(2006)}]{KitaevPreskillTEE}%
  \BibitemOpen
  \bibfield  {author} {\bibinfo {author} {\bibfnamefont {Alexei}\ \bibnamefont {{Kitaev}}}\ and\ \bibinfo {author} {\bibfnamefont {John}\ \bibnamefont {{Preskill}}},\ }\bibfield  {title} {\enquote {\bibinfo {title} {{Topological Entanglement Entropy}},}\ }\href {\doibase 10.1103/PhysRevLett.96.110404} {\bibfield  {journal} {\bibinfo  {journal} {\prl}\ }\textbf {\bibinfo {volume} {96}},\ \bibinfo {eid} {110404} (\bibinfo {year} {2006})},\ \Eprint {http://arxiv.org/abs/hep-th/0510092} {arXiv:hep-th/0510092 [hep-th]} \BibitemShut {NoStop}%
\bibitem [{\citenamefont {{Levin}}\ and\ \citenamefont {{Wen}}(2006)}]{LevinWenTEE}%
  \BibitemOpen
  \bibfield  {author} {\bibinfo {author} {\bibfnamefont {Michael}\ \bibnamefont {{Levin}}}\ and\ \bibinfo {author} {\bibfnamefont {Xiao-Gang}\ \bibnamefont {{Wen}}},\ }\bibfield  {title} {\enquote {\bibinfo {title} {{Detecting Topological Order in a Ground State Wave Function}},}\ }\href {\doibase 10.1103/PhysRevLett.96.110405} {\bibfield  {journal} {\bibinfo  {journal} {\prl}\ }\textbf {\bibinfo {volume} {96}},\ \bibinfo {eid} {110405} (\bibinfo {year} {2006})},\ \Eprint {http://arxiv.org/abs/cond-mat/0510613} {arXiv:cond-mat/0510613 [cond-mat.str-el]} \BibitemShut {NoStop}%
\bibitem [{\citenamefont {{Potter}}\ \emph {et~al.}(2017)\citenamefont {{Potter}}, \citenamefont {{Wang}}, \citenamefont {{Metlitski}},\ and\ \citenamefont {{Vishwanath}}}]{Potter2017}%
  \BibitemOpen
  \bibfield  {author} {\bibinfo {author} {\bibfnamefont {Andrew~C.}\ \bibnamefont {{Potter}}}, \bibinfo {author} {\bibfnamefont {Chong}\ \bibnamefont {{Wang}}}, \bibinfo {author} {\bibfnamefont {Max~A.}\ \bibnamefont {{Metlitski}}}, \ and\ \bibinfo {author} {\bibfnamefont {Ashvin}\ \bibnamefont {{Vishwanath}}},\ }\bibfield  {title} {\enquote {\bibinfo {title} {{Realizing topological surface states in a lower-dimensional flat band}},}\ }\href {\doibase 10.1103/PhysRevB.96.235114} {\bibfield  {journal} {\bibinfo  {journal} {\prb}\ }\textbf {\bibinfo {volume} {96}},\ \bibinfo {eid} {235114} (\bibinfo {year} {2017})},\ \Eprint {http://arxiv.org/abs/1609.08618} {arXiv:1609.08618 [cond-mat.str-el]} \BibitemShut {NoStop}%
\bibitem [{\citenamefont {Essin}\ and\ \citenamefont {Hermele}(2013)}]{hermeleSET}%
  \BibitemOpen
  \bibfield  {author} {\bibinfo {author} {\bibfnamefont {Andrew~M.}\ \bibnamefont {Essin}}\ and\ \bibinfo {author} {\bibfnamefont {Michael}\ \bibnamefont {Hermele}},\ }\bibfield  {title} {\enquote {\bibinfo {title} {Classifying fractionalization: Symmetry classification of gapped ${\mathbb{z}}_{2}$ spin liquids in two dimensions},}\ }\href {\doibase 10.1103/PhysRevB.87.104406} {\bibfield  {journal} {\bibinfo  {journal} {Phys. Rev. B}\ }\textbf {\bibinfo {volume} {87}},\ \bibinfo {pages} {104406} (\bibinfo {year} {2013})}\BibitemShut {NoStop}%
\bibitem [{\citenamefont {Mesaros}\ and\ \citenamefont {Ran}(2013)}]{ranSET}%
  \BibitemOpen
  \bibfield  {author} {\bibinfo {author} {\bibfnamefont {Andrej}\ \bibnamefont {Mesaros}}\ and\ \bibinfo {author} {\bibfnamefont {Ying}\ \bibnamefont {Ran}},\ }\bibfield  {title} {\enquote {\bibinfo {title} {Classification of symmetry enriched topological phases with exactly solvable models},}\ }\href {\doibase 10.1103/PhysRevB.87.155115} {\bibfield  {journal} {\bibinfo  {journal} {Phys. Rev. B}\ }\textbf {\bibinfo {volume} {87}},\ \bibinfo {pages} {155115} (\bibinfo {year} {2013})}\BibitemShut {NoStop}%
\bibitem [{\citenamefont {Hung}\ and\ \citenamefont {Wen}(2013)}]{wenSET}%
  \BibitemOpen
  \bibfield  {author} {\bibinfo {author} {\bibfnamefont {Ling-Yan}\ \bibnamefont {Hung}}\ and\ \bibinfo {author} {\bibfnamefont {Xiao-Gang}\ \bibnamefont {Wen}},\ }\bibfield  {title} {\enquote {\bibinfo {title} {Quantized topological terms in weak-coupling gauge theories with a global symmetry and their connection to symmetry-enriched topological phases},}\ }\href {\doibase 10.1103/PhysRevB.87.165107} {\bibfield  {journal} {\bibinfo  {journal} {Phys. Rev. B}\ }\textbf {\bibinfo {volume} {87}},\ \bibinfo {pages} {165107} (\bibinfo {year} {2013})}\BibitemShut {NoStop}%
\bibitem [{\citenamefont {Lu}\ and\ \citenamefont {Vishwanath}(2016)}]{luSET}%
  \BibitemOpen
  \bibfield  {author} {\bibinfo {author} {\bibfnamefont {Yuan-Ming}\ \bibnamefont {Lu}}\ and\ \bibinfo {author} {\bibfnamefont {Ashvin}\ \bibnamefont {Vishwanath}},\ }\bibfield  {title} {\enquote {\bibinfo {title} {Classification and properties of symmetry-enriched topological phases: Chern-simons approach with applications to ${Z}_{2}$ spin liquids},}\ }\href {\doibase 10.1103/PhysRevB.93.155121} {\bibfield  {journal} {\bibinfo  {journal} {Phys. Rev. B}\ }\textbf {\bibinfo {volume} {93}},\ \bibinfo {pages} {155121} (\bibinfo {year} {2016})}\BibitemShut {NoStop}%
\bibitem [{\citenamefont {Chen}\ \emph {et~al.}(2013)\citenamefont {Chen}, \citenamefont {Gu}, \citenamefont {Liu},\ and\ \citenamefont {Wen}}]{wenspt1}%
  \BibitemOpen
  \bibfield  {author} {\bibinfo {author} {\bibfnamefont {Xie}\ \bibnamefont {Chen}}, \bibinfo {author} {\bibfnamefont {Zheng-Cheng}\ \bibnamefont {Gu}}, \bibinfo {author} {\bibfnamefont {Zheng-Xin}\ \bibnamefont {Liu}}, \ and\ \bibinfo {author} {\bibfnamefont {Xiao-Gang}\ \bibnamefont {Wen}},\ }\bibfield  {title} {\enquote {\bibinfo {title} {Symmetry protected topological orders and the group cohomology of their symmetry group},}\ }\href {\doibase 10.1103/PhysRevB.87.155114} {\bibfield  {journal} {\bibinfo  {journal} {Phys. Rev. B}\ }\textbf {\bibinfo {volume} {87}},\ \bibinfo {pages} {155114} (\bibinfo {year} {2013})}\BibitemShut {NoStop}%
\bibitem [{\citenamefont {Chen}\ \emph {et~al.}(2012)\citenamefont {Chen}, \citenamefont {Gu}, \citenamefont {Liu},\ and\ \citenamefont {Wen}}]{wenspt2}%
  \BibitemOpen
  \bibfield  {author} {\bibinfo {author} {\bibfnamefont {Xie}\ \bibnamefont {Chen}}, \bibinfo {author} {\bibfnamefont {Zheng-Cheng}\ \bibnamefont {Gu}}, \bibinfo {author} {\bibfnamefont {Zheng-Xin}\ \bibnamefont {Liu}}, \ and\ \bibinfo {author} {\bibfnamefont {Xiao-Gang}\ \bibnamefont {Wen}},\ }\href@noop {} {\bibfield  {journal} {\bibinfo  {journal} {Science}\ }\textbf {\bibinfo {volume} {338}},\ \bibinfo {pages} {1604} (\bibinfo {year} {2012})}\BibitemShut {NoStop}%
\bibitem [{\citenamefont {Senthil}\ and\ \citenamefont {Levin}(2013)}]{levinsenthil}%
  \BibitemOpen
  \bibfield  {author} {\bibinfo {author} {\bibfnamefont {T.}~\bibnamefont {Senthil}}\ and\ \bibinfo {author} {\bibfnamefont {Michael}\ \bibnamefont {Levin}},\ }\bibfield  {title} {\enquote {\bibinfo {title} {Integer quantum hall effect for bosons},}\ }\href {\doibase 10.1103/PhysRevLett.110.046801} {\bibfield  {journal} {\bibinfo  {journal} {Phys. Rev. Lett.}\ }\textbf {\bibinfo {volume} {110}},\ \bibinfo {pages} {046801} (\bibinfo {year} {2013})}\BibitemShut {NoStop}%
\bibitem [{\citenamefont {Lu}\ and\ \citenamefont {Vishwanath}(2012)}]{luashvin}%
  \BibitemOpen
  \bibfield  {author} {\bibinfo {author} {\bibfnamefont {Yuan-Ming}\ \bibnamefont {Lu}}\ and\ \bibinfo {author} {\bibfnamefont {Ashvin}\ \bibnamefont {Vishwanath}},\ }\bibfield  {title} {\enquote {\bibinfo {title} {Theory and classification of interacting integer topological phases in two dimensions: A chern-simons approach},}\ }\href {\doibase 10.1103/PhysRevB.86.125119} {\bibfield  {journal} {\bibinfo  {journal} {Phys. Rev. B}\ }\textbf {\bibinfo {volume} {86}},\ \bibinfo {pages} {125119} (\bibinfo {year} {2012})}\BibitemShut {NoStop}%
\bibitem [{\citenamefont {Xu}\ and\ \citenamefont {Senthil}(2013)}]{xusenthil}%
  \BibitemOpen
  \bibfield  {author} {\bibinfo {author} {\bibfnamefont {Cenke}\ \bibnamefont {Xu}}\ and\ \bibinfo {author} {\bibfnamefont {T.}~\bibnamefont {Senthil}},\ }\bibfield  {title} {\enquote {\bibinfo {title} {Wave functions of bosonic symmetry protected topological phases},}\ }\href {\doibase 10.1103/PhysRevB.87.174412} {\bibfield  {journal} {\bibinfo  {journal} {Phys. Rev. B}\ }\textbf {\bibinfo {volume} {87}},\ \bibinfo {pages} {174412} (\bibinfo {year} {2013})}\BibitemShut {NoStop}%
\bibitem [{Note2()}]{Note2}%
  \BibitemOpen
  \bibinfo {note} {$n_{1,2}$ should be viewed as integers mod $8$}\BibitemShut {NoStop}%
\bibitem [{\citenamefont {Xia}\ \emph {et~al.}(2024)\citenamefont {Xia}, \citenamefont {Han}, \citenamefont {Watanabe}, \citenamefont {Taniguchi}, \citenamefont {Shan},\ and\ \citenamefont {Mak}}]{TMDSC}%
  \BibitemOpen
  \bibfield  {author} {\bibinfo {author} {\bibfnamefont {Yiyu}\ \bibnamefont {Xia}}, \bibinfo {author} {\bibfnamefont {Zhongdong}\ \bibnamefont {Han}}, \bibinfo {author} {\bibfnamefont {Kenji}\ \bibnamefont {Watanabe}}, \bibinfo {author} {\bibfnamefont {Takashi}\ \bibnamefont {Taniguchi}}, \bibinfo {author} {\bibfnamefont {Jie}\ \bibnamefont {Shan}}, \ and\ \bibinfo {author} {\bibfnamefont {Kin~Fai}\ \bibnamefont {Mak}},\ }\href@noop {} {\enquote {\bibinfo {title} {Unconventional superconductivity in twisted bilayer wse2},}\ } (\bibinfo {year} {2024}),\ \Eprint {http://arxiv.org/abs/2405.14784} {arXiv:2405.14784 [cond-mat.mes-hall]} \BibitemShut {NoStop}%
\bibitem [{\citenamefont {Guo}\ \emph {et~al.}(2024)\citenamefont {Guo}, \citenamefont {Pack}, \citenamefont {Swann}, \citenamefont {Holtzman}, \citenamefont {Cothrine}, \citenamefont {Watanabe}, \citenamefont {Taniguchi}, \citenamefont {Mandrus}, \citenamefont {Barmak}, \citenamefont {Hone}, \citenamefont {Millis}, \citenamefont {Pasupathy},\ and\ \citenamefont {Dean}}]{TMDSC2}%
  \BibitemOpen
  \bibfield  {author} {\bibinfo {author} {\bibfnamefont {Yinjie}\ \bibnamefont {Guo}}, \bibinfo {author} {\bibfnamefont {Jordan}\ \bibnamefont {Pack}}, \bibinfo {author} {\bibfnamefont {Joshua}\ \bibnamefont {Swann}}, \bibinfo {author} {\bibfnamefont {Luke}\ \bibnamefont {Holtzman}}, \bibinfo {author} {\bibfnamefont {Matthew}\ \bibnamefont {Cothrine}}, \bibinfo {author} {\bibfnamefont {Kenji}\ \bibnamefont {Watanabe}}, \bibinfo {author} {\bibfnamefont {Takashi}\ \bibnamefont {Taniguchi}}, \bibinfo {author} {\bibfnamefont {David}\ \bibnamefont {Mandrus}}, \bibinfo {author} {\bibfnamefont {Katayun}\ \bibnamefont {Barmak}}, \bibinfo {author} {\bibfnamefont {James}\ \bibnamefont {Hone}}, \bibinfo {author} {\bibfnamefont {Andrew~J.}\ \bibnamefont {Millis}}, \bibinfo {author} {\bibfnamefont {Abhay~N.}\ \bibnamefont {Pasupathy}}, \ and\ \bibinfo {author} {\bibfnamefont {Cory~R.}\ \bibnamefont {Dean}},\ }\href@noop {} {\enquote {\bibinfo {title} {Superconductivity in twisted bilayer wse$_2$},}\ } (\bibinfo {year}
  {2024}),\ \Eprint {http://arxiv.org/abs/2406.03418} {arXiv:2406.03418 [cond-mat.mes-hall]} \BibitemShut {NoStop}%
\bibitem [{\citenamefont {Devakul}\ \emph {et~al.}(2021{\natexlab{b}})\citenamefont {Devakul}, \citenamefont {Crépel}, \citenamefont {Zhang},\ and\ \citenamefont {Fu}}]{Devakul_2021}%
  \BibitemOpen
  \bibfield  {author} {\bibinfo {author} {\bibfnamefont {Trithep}\ \bibnamefont {Devakul}}, \bibinfo {author} {\bibfnamefont {Valentin}\ \bibnamefont {Crépel}}, \bibinfo {author} {\bibfnamefont {Yang}\ \bibnamefont {Zhang}}, \ and\ \bibinfo {author} {\bibfnamefont {Liang}\ \bibnamefont {Fu}},\ }\bibfield  {title} {\enquote {\bibinfo {title} {Magic in twisted transition metal dichalcogenide bilayers},}\ }\href {\doibase 10.1038/s41467-021-27042-9} {\bibfield  {journal} {\bibinfo  {journal} {Nature Communications}\ }\textbf {\bibinfo {volume} {12}} (\bibinfo {year} {2021}{\natexlab{b}}),\ 10.1038/s41467-021-27042-9}\BibitemShut {NoStop}%
\bibitem [{\citenamefont {May-Mann}\ \emph {et~al.}(2024)\citenamefont {May-Mann}, \citenamefont {Stern},\ and\ \citenamefont {Devakul}}]{Julian}%
  \BibitemOpen
  \bibfield  {author} {\bibinfo {author} {\bibfnamefont {Julian}\ \bibnamefont {May-Mann}}, \bibinfo {author} {\bibfnamefont {Ady}\ \bibnamefont {Stern}}, \ and\ \bibinfo {author} {\bibfnamefont {Trithep}\ \bibnamefont {Devakul}},\ }\href@noop {} {\enquote {\bibinfo {title} {Theory of half-integer fractional quantum spin hall insulator edges},}\ } (\bibinfo {year} {2024}),\ \Eprint {http://arxiv.org/abs/2403.03964} {arXiv:2403.03964 [cond-mat.mes-hall]} \BibitemShut {NoStop}%
\bibitem [{\citenamefont {{Zhang}}(2024{\natexlab{b}})}]{ZhangFQSH}%
  \BibitemOpen
  \bibfield  {author} {\bibinfo {author} {\bibfnamefont {Ya-Hui}\ \bibnamefont {{Zhang}}},\ }\bibfield  {title} {\enquote {\bibinfo {title} {{Non-Abelian and Abelian descendants of vortex spin liquid: fractional quantum spin Hall effect in twisted MoTe$_2$}},}\ }\href {\doibase 10.48550/arXiv.2403.12126} {\bibfield  {journal} {\bibinfo  {journal} {arXiv e-prints}\ ,\ \bibinfo {eid} {arXiv:2403.12126}} (\bibinfo {year} {2024}{\natexlab{b}})},\ \Eprint {http://arxiv.org/abs/2403.12126} {arXiv:2403.12126 [cond-mat.str-el]} \BibitemShut {NoStop}%
\bibitem [{\citenamefont {Müger}(2003)}]{Muger_2003}%
  \BibitemOpen
  \bibfield  {author} {\bibinfo {author} {\bibfnamefont {Michael}\ \bibnamefont {Müger}},\ }\bibfield  {title} {\enquote {\bibinfo {title} {On the structure of modular categories},}\ }\href {\doibase 10.1112/S0024611503014187} {\bibfield  {journal} {\bibinfo  {journal} {Proceedings of the London Mathematical Society}\ }\textbf {\bibinfo {volume} {87}},\ \bibinfo {pages} {291–308} (\bibinfo {year} {2003})}\BibitemShut {NoStop}%
\bibitem [{\citenamefont {{Lapa}}\ and\ \citenamefont {{Levin}}(2019)}]{Lapa2019Indicator}%
  \BibitemOpen
  \bibfield  {author} {\bibinfo {author} {\bibfnamefont {Matthew~F.}\ \bibnamefont {{Lapa}}}\ and\ \bibinfo {author} {\bibfnamefont {Michael}\ \bibnamefont {{Levin}}},\ }\bibfield  {title} {\enquote {\bibinfo {title} {{Anomaly indicators for topological orders with U (1 ) and time-reversal symmetry}},}\ }\href {\doibase 10.1103/PhysRevB.100.165129} {\bibfield  {journal} {\bibinfo  {journal} {\prb}\ }\textbf {\bibinfo {volume} {100}},\ \bibinfo {eid} {165129} (\bibinfo {year} {2019})},\ \Eprint {http://arxiv.org/abs/1905.00435} {arXiv:1905.00435 [cond-mat.str-el]} \BibitemShut {NoStop}%
\end{thebibliography}%

\end{document}